\documentclass[10pt,a4paper]{article}
\usepackage[utf8]{inputenc}
\usepackage{graphicx}
\usepackage{authblk}
\usepackage{amsmath}

\usepackage{caption}
\usepackage{subcaption}

\title{Dynamics of dipolar Atom-Molecular BEC in a double well potential: Effect of atom-molecular coherent coupling}

\author[1,*]{Krishna Rai Dastidar}
\author[2]{ Moumita Gupta}
\affil[1]{Indian Association for the Cultivation of Science, Kolkata-700032, INDIA}
\affil[2] {Charuchandra College, Kolkata-700029, INDIA}

\date{}

\begin{document}

\maketitle
* spkrd@iacs.res.in

\begin{abstract}
In the present work we have studied the dynamics of dipolar
atom-molecular Bose Einstein Condensates coupled via Feshbach
Resonance in a double well potential. We have numerically solved
four coupled GP like equations, two for left well and two for
right well for this atom molecular coupled system. Our numerical
results show that both the long-range dipole-dipole interaction
(chosen to be positive) and the coherent coupling interaction
(which is positive for bosons) facilitate the transmission of
atoms and molecules from left well to the right well when the
population in the right well dominates over that in the left well
and is trapped for a period of time. Whereas in absence of any one
of these interactions probability of  transient transmission
decreases. However in absence of both the interactions
(dipole-dipole and coherent coupling) i.e. when only the
repulsive contact interaction is present, it leads to self
trapping in the left well for a period of time.  It is also shown
that the signature of coherent coupling between atoms and
molecules on the density distribution of atoms in the double well
potential is present both in absence and presence of dipole-dipole
interaction.
\end{abstract}

\section{Introduction}
The experimental observation of Bose-Einstein condensation in a
dilute trapped gas of alkali-metal atoms \cite{Anderson1995} has
stimulated a number of theoretical and experimental studies on the
properties of these condensates. One of these fields is the study
of steady state and dynamical properties  of these condensates
while trapped in a double well or multi well configurations
\cite{Holthas2001, Lahaye2009}. It explored the possibility of
realising Josephson junction in ultra-cold atoms or condensates.
Josephson effect (JE) in ultra-cold atoms was first predicted by
Javanainen \cite{Javanainen1986} and theoretically realized in a
double well potential occupied by a macroscopic number of bosons
\cite{Grossmann1995}. It has also been shown that coherent quantum
tunneling between two condensates in two traps gives rise to
oscillation in atomic numbers \cite{Jack1996}. The Josephson
oscillation (JO) in atomic condensate in a double well potential
was studied including the many body hard sphere interaction and
discussed the validity of two-mode model for JO
\cite{Milburn1997}. Besides the Josephson oscillations in atomic
condensates in double well potential a new phenomena i.e.
macroscopic quantum self-trapping (MQST) can be obtained which has
no analog for superconducing Josephson junctions. MQST was first
predicted by Smerzi and collaborators \cite{Smerzi1997} and
subsequently both the Josephson oscillation (JO) and MQST were
observed in single bosonic Josephson junction (BJJ)
\cite{Albiez2005, Gati2007}. JE was also observed in an array of
BJJ/periodic potential \cite{Anker2005, Cataliotti2001} and
theoretical investigation was carried out to provide an
explanation of the observations in periodic potential
\cite{Alexandar2006, Wang2006}. JO and MQST has been studied also
in triple well \cite{Liu2007}. Ananikian and Bergeman
\cite{Ananikian2006} have studied validity of two-mode model for
BECs in double well potential and by using improved two-mode model
reproduced the experimentally observed \cite{Albiez2005} tunneling
oscillation frequency. Julia-DIaz et al \cite{Julia2010} and
Mele-Messeguer et al \cite{Messeguer2011} studied JO in binary
BECs in double well potential and explored the dependence of JO on
different spin collision channels. Study on self trapping of BEC
in DW shows  by using periodically modulated double well potential
MQST can be systematically controlled. Li-Hua Lu and You-Quan Li
\cite{Lu2009} has shown that in partially coherent BEC in DW  both
degree of coherence and initial phase difference affects the final
particle distribution in two wells. Ottaviani et al \cite
{Ottaviani2010} studied transport and self-trapping of BEC in a
double well by controlling the energy bias between two wells as
well as nonlinearity of the system. Shchesnovich and Trippenbach
\cite {Trippenbach2008} used Fock space method to study JO and
MQST in BEC in a double well potential and suggested that their
model can also be applied for atom-molecular coherence and
nonlinear tunneling of BEC in optical lattices.
 Nonlinear Josephson dynamics of BECs in a double-well potential has been studied
 including coupling to quasiparticle states and it has been suggested that JO and Rabi oscillation
 can be damped due to quasiparticle inelastic collisions \cite{Martinez2009}.
Effect of decoherence on the Josephson dynamics of BEC in DW
potential has been studied by analyzing the coupling of the
condensate with the environment and have shown that damping of JO
due to decoherence can enhance the self-trapping \cite{Wang2007}.

 Time-dependent self-trapping of BECs in a DW
potential has been discussed by phase space analysis within the
mean field approximation and have shown that the population
imbalance of BECs can be controlled by varying the atom-atom
interaction and the driving time dependent magnetic field
\cite{Cui2010}. Gillet et al. \cite{Gillet2014} have studied
Josephson dynamics in BECs in a symmetric one dimensional double
well potential within the four mode approximation and have shown
that the dissipation due to the presence of two excited modes
affects the rich Josephson dynamics in BECs. Huang et al
\cite{Huang2012} have used quantum Fisher information to study
evolution of JO to MQST and the parameter sensitivity of the
stability of the condensate system. L. Fu and J. Liu \cite{Fu2006}
discussed the influence of many-body quantum fluctuations on the
transition to self-trapping in a BEC in a symmetric double well.
They showed manifestation of quantum entanglement  of the
transition which reaches  to maximum at the transition point. Levy
et al. \cite{Levy2007} studied dc and ac Josephson effects in two
weakly linked Bose-Einstein condensates considering a
time-dependent barrier, moving adiabatically across the trapping
potential and have shown that there is a critical velocity of the
barrier  at which sharp transition from dc to ac JE occurs. The
transport of Bose-Einstein condensate (BEC) in a double-well trap
can be controlled by considering the time-dependent (Gaussian)
coupling between the wells and their relative detuning within the
mean field approximation both for repulsive and attractive
interactions \cite{Nesterenko2012}.  This group has also explored
the inverse population transfer of the repulsive Bose-Einstein
condensate (BEC) in a weakly bound double-well trap within the 3D
time-dependent Gross-Pitaevskii equation considering a time
dependent barrier shift separating the two wells.

Josephson effects such as JO and MQST have also been studied in
bosonic and fermonic mixtures in double well potential
\cite{Adhikari2009}. Luick et al \cite{Luick2020} measured the
frequency of Josephson oscillations as a function of the phase
difference across the junction and found excellent agreement with
the sinusoidal current phase relation in a strongly correlated 2D
fermi gas. In our group we consider bosons or fermions trapped in
a quasi-1D DW potential interacting via a 3D finite range two-body
interaction potential with large scattering length $a_s$. Under
tight-binding or two mode approximation, we describe in detail the
effects of the range of interaction on the quantum dynamics and
number-phase uncertainty in the strongly interacting or unitarity
regime. We defined the standard quantum limit (SQL) for phase and
number fluctuations and described two-mode squeezing for number
and phase variables for this system \cite{Adhikary2018}.

With the observation of BECs of dipolar atoms with large magnetic
moments e.g. $^{52}Cr$, $^{168}Er$, $^{164}Dy$ etc
\cite{Griesmaier2005} stimulated a large number of theoretical and
experimental investigations on the effect of long range
dipole-dipole interaction on the properties of BECs. The presence
of long range, anisotropic and nonlocal dipole-dipole interactions
the manifestation of JE in dipolar BECs becomes different than
that in case of short range contact interactions. The dipolar
interaction may drastically change the properties of quasi-1D
dipolar condensates, even for situations in which the dipolar
interaction would be completely overwhelmed by the short-range
interactions in a 3D environment \cite{Santos2007}. With the
experimental advent in trapping and cooling of dipolar molecular
gases of both fermionic and bosonic nature has stimulated the
study of the properties of dipolar BECs e.g. structure, collective
excitations, chemical reaction between dipolar molecules, vortex
in spin-orbit coupled rotating dipolar molecules, dipolar BEC and
superfluid BCS pairing transition etc \cite{Ni2008}. The stability
of dipolar BECs of $^{52}Cr$ has been experimentally investigated
which depends on the scattering length and strongly on the trap
geometry \cite{Koch2008}. First realization of heteronuclear
dipolar quantum mixtures of highly magnetic erbium and dysprosium
atoms has been reported and experimentally demonstrated binary
Bose-Einstein condensation in five different Er-Dy isotope
combinations, as well as one Er-Dy Bose-Fermi mixture
\cite{Trautmann2018}.

Mixing, demixing, and structure formation in a binary dipolar
Bose-Einstein condensate \cite{Adhikari2012} and the miscibility
in coupled dipolar and non-dipolar Bose-Einstein condensates
 have been studied. Stability of three-dimensional
vortex lattice structures in purely dipolar Bose-Einstein
condensate (BEC) has been experimentally and theoretically studied
by varying the aspect ratio and the strength of dipole-dipole
interaction \cite{Kumar2016}. Vortex lattice formation in dipolar
BECs via rotation of the polarisation and the orientation of
dipoles has also been explored \cite{Prasad2019, Yi2006}.
Anisotropic superfluid flow has been observed in dipolar BEC with
strong magnetic dipole-dipole interaction by moving an attractive
laser beam through the condensate \cite{Wenzel2018}. Formation of
spin-vortex pair, solitons, vortex-solitons  and its stability in
dipolar spinor BECs have been examined \cite{Zhang2016}.
 Within the Hamiltonian mean-field
(HMF) model it has been shown that the stationary solutions of
generalized Gross-Pitaevskii equation (GGPE) can give rise to a
tower of solitons in presence of long range dipolar interactions
in a BEC \cite{Plestid2019}. To study structure and dynamics of
trapped dipolar gases different theoretical models e.g. two
mode/multimode, mean field approximation, numerical integration of
Gross-Pitaevskii/modified Gross-Pitaevskii  equation etc have been
used. However we have derived analytical solutions for
Gross-Pitaevskii and modified Gross-Pitaevskii  equations for
non-dipolar atomic condensate trapped in harmonic potential using
modified Thomas-Fermi approximation \cite {Gupta2008} and
 Eberlien et al \cite{Eberlein2005} first derived exact solution of the
Thomas-Fermi equation for a trapped Bose-Einstein condensate with
dipole-dipole interactions.

One of the interesting features in
dipolar BECs is the appearance of Roton minimum in the spectrum of
elementary excitations \cite{Santos2003}  which occurs when the
dipole-dipole long range interaction dominates over short range
contact interaction. Roton minimum in dipolar ultra cold gases has
been observed  and theoretical analysed
\cite{Chomaz2018,Petter2019}. Instabiliities in rotonization of 1D
dipolar gas in 1D lattices and also in 2D dipolar gases have been
discussed \cite{Giovanazzi2004}. Softening of Roton mode has also been studied by
considering finite range interaction and laser induced
dipole-dipole interactions in  single and multi-component BECs.
Effect of tilting of the dipoles on the Roton mode excitation and
post-roton instability in a dipolar BEC confined in
pancake-shaped trap has been studied \cite{Mishra2016}. In the region of post-roton
instability in dipolar BECs it is demonstrated both theoretically
and experimentally stable droplets can be formed and it is
stabilised by the quantum fluctuation/three-body interaction
\cite{Adhikari2017}. Moreover
quantum fluctuations in dipolar condensates stabilize a new
supersolid phase giving rise to high modulational contrast in
density(cristalization) and near-perfect superfluidity
\cite{Zhang2019, Kora2019}. Supersolidity of dipolar BEC has been
extensively studied in recent experiments and theoretical
explanation was also provided \cite{Tanzi2019}.
  The advances of
theoretical and experimental studies in dipolar quantum gases has
been reviewed and the physical insight has been presented by
Lahaye et al\cite{Lahaye2009}.

In general in the study of JE in non-dipolar BECs in DW or multi
well potentials nonlinear short range  contact interactions are
considered which controls the transition from JO to MQST. However
for dipolar BECs in DW/multi well potentials interplay between
short range contact interaction and the long-range dipole-dipole
interaction leading to switch over from JO to MQST and vice versa
has been studied by varying the scattering length and the number
of atoms in dipolar BEC trapped in double well/optical lattice. It
is shown that the system can translate from MQST to JO in presence
of  dipolar interaction \cite{Adhikari2014}. Coherent tunneling
phenomena leading to JO and MQST can occur in an effective
ring-shaped double well potential induced by the dipolar
interaction \cite{Abad2011, Polo2019}.  Dependence of JE on
alignment of dipoles/ magnetisation direction and the stability of
the dipolar BEC in double well have been theoretically
demonstrated \cite{Asad2009}. Effect of excitation to higher
excited state on the dynamical stabilisation of dipolar BEC in
triple well configuration has been discussed \cite{Fortanier2013}.
It has also been shown introduction of asymmetric basis function
can improve the results from two-mode model \cite{Wang2020}. Spin
dynamics of ultra cold magnetic atoms located in two wells has
been experimentally studied and the effect of long-range dipolar
interactions on the dynamics has been also explored theoretically
\cite{Paz2014, Ashhab2002}. Recently our group has studied the
effect of trap confinement and the atom-atom nonlinear interaction
on JO and MQST in cold atoms in a DW potential. We considered
three types of interaction potentials (i) contact, (ii)long range
dipolar and (iii) finite range potentials. We found that by
varying the aspect ratio of the trap (i) transition from JO to
MQSt occurs  for small atom-atom interactions (ii) transition from
Rabi to JO and JO to MQST occurs for long range dipolar
interaction and transition from JO to MQST occurs even if
scattering length is relatively large in the region of narrow
Feshbach resonance due to finite range effects \cite {Saha2019}.

Besides Josephson effect, formation of
vortices/vortex solitons in dipolar BECs confined in double
well/optical lattice has been explored \cite{Subramaniyan2017}.
Formation of hidden vortices within the barrier and
vortex lattice has been demonstrated in dipolar BEC in a double
well potential \cite{Wen2010}.

With the advent of experimental realization of state selective
molecules from atomic BECs/ultracold atoms by photo association
(PA) \cite{Wynar2000, Gerton2000}, by magnetic Feshbach resonance
(FR)\cite{Donley2002} and by  magneto-photoaasociation applying
Feshbach resonance (FR) technique \cite{Winkler2007},  many
theoretical attempts were made to understand and set the
guidelines for the realization of molecular BECs by the process of
PA and magnetic FR \cite{Heinzen2000}. In our group an attempt was
made \cite {Gupta2009} to explain the experimental observation
\cite {Donley2002} on the dynamics of atom-molecular coherence in
BEC of $^{85}Rb$ atoms. Bose-Einstein condensation of $Li_2$
molecules from  fermionic Lithium atoms in an optical trap was
reported \cite{Jochim2003}. Dynamics of coupled atom-molecular
condensates in presence of PA in a double well potential was
studied to explore the effect  of tunneling and PA laser intensity
on the number of molecules formed \cite{Jing2006}. Stability and
JE of atom molecular coupled BECs in a double well potential has
also been studied \cite{Atsushi2009} by considering the four mode
model.

In this paper we have studied dynamics of dipolar atomic
condensates coupled with  molecular condensate via magnetic
Feshbach resonance confined in a double well potential. We
explored the effect of atom-molecular coherence and the effect of
long-range dipole-dipole interactions on the transport of atoms
and molecules from one well to the other. In this theoretical
approach we solved four time dependent coupled Gross-Pitaevskii
equations (two for each well) to investigate the temporal
evolution of atomic and molecular numbers in each well. To our
knowledge the effect of atom-molecular coherence  on the dynamics
of dipolar atomic condensates in double well potential, coupled
with molecules via magnetic Feshbach resonance has not been
studied yet.

\section {Theory}
In the present work we have studied numerically the dynamics of
coupled dipolar atom-molecular BECs from left to right well and
vice versa in a double  well potential in the axial direction.
Atomic and molecular BECs are coupled via magnetic Feshbach
resonance \cite {Donley2002, Gupta2009} in both the wells. Four
coupled time-dependent GP like equations for atomic and molecular
BECs two for left  and two for right well have been derived from
the first principle considering the mean field energy of the
coupled system. Four time-dependent three dimensional (3D) coupled
GP like equations were then reduced to single dimension (1D) to
study the dynamics of dipolar coupled BECs trapped in a double
well potential in the axial direction.

\subsection{\bf {Derivation of four coupled equations}:}
Mean field energy density for dipolar atom-molecular coupled condensate in double well is given as:
$$ E[ \psi_{al}, \psi_{ar}, \psi_{ml},\psi_{mr}]=
\psi_{al}^*[-\frac{\hbar^2\nabla^2}{2m}+v_a({\bf r})+ \frac{\lambda_a}
{2}  {\psi_{al}^*\psi_{al}}+$$$$\frac{d_a}{2}{\int {V ({\bf
r}-{\bf r'})} \psi_{al}^* \psi_{al} d{\bf r'}}]\psi_{al}
+\psi_{ar}^*[-\frac{\hbar^2\nabla^2}{
2m}+v_a({\bf r})+\frac{\lambda_a}{2} \psi_{ar}^*\psi_{ar}+$$$$\frac{d_a}{
2} \int {V ({\bf r}-{\bf r'}) \psi_{ar}^* \psi_{ar} d{\bf
r'}}]\psi_{ar}+\psi_{ml}^*[-\frac{\hbar^2\nabla^2}{
4m}+v_m({\bf r})+$$$$\frac{\lambda_m}{2}
\psi_{ml}^*\psi_{ml}+\epsilon+\frac{d_m}{2} \int {V ({\bf r}-{\bf
r'}) \psi_{ml}^* \psi_{ml} d{\bf r'}}]\psi_{ml}+$$
$$\psi_{mr}^*[-\frac{\hbar^2\nabla^2}{
4m}+v_m({\bf r})+\frac{\lambda_m}{2}
\psi_{mr}^*\psi_{mr}+\epsilon+\frac{d_m}{2}\int {V ({\bf r}-{\bf
r'}) \psi_{mr}^* \psi_{mr} d{\bf r'}}]\psi_{mr}+$$
$$\lambda_{am}[\psi_{al}^* \psi_{al} \psi_{ml}^* \psi_{ml}+
\psi_{ar}^* \psi_{ar} \psi_{mr}^* \psi_{mr}]+$$
\begin{eqnarray}
\frac{\chi}{2}[ \psi_{mr} \psi_{ar}^* \psi_{ar}^*+
 \psi_{ml} \psi_{al}^* \psi_{al}^*]+
 \frac{\chi}{2}[\psi_{ml}^*\psi_{al}\psi_{al}+
\psi_{mr}^*\psi_{ar}\psi_{ar}]
\end{eqnarray}
Here $(\psi_{al}, \psi_{ar})$ are the macroscopic wavefunctions
for atomic condensate in left and right wells respectively and
$(\psi_{ml}, \psi_{mr})$ are the same for the molecular condensate
in the double well respectively. All the wavefunctions mentioned
above are the function of $ {\bf r}$ and t. $v_a({\bf r})$ and
$v_m({\bf r})$ are the double well potential for atomic and
molecular traps respectively, which are given as: $$v_a({\bf
r})={1\over 2} {m\omega_z^2({\rho^2\over{\lambda^2}}+z^2)}+A
e^{-\kappa z^2}$$ and
$$v_m({\bf r})= {m\omega_z^2({\rho^2\over{\lambda^2}}+z^2)}+A e^{-\kappa z^2}$$ where $\omega_\rho,$ $\omega_z$ are the radial and axial trap frequencies and
the trap asymmetry parameter $\lambda= {\omega_z\over{\omega_\rho}}$. The constants A and $\kappa$ are the height and the width of the Gaussian barrier which gives
rise to double well feature in the axial direction when added to the harmonic potential.
$\lambda_a, \lambda_m$ and $\lambda_{am}$ are the s-wave contact interactions between atom-atom, molecule-molecule and atom-molecule respectively.
$\lambda_a={4\pi\hbar^2 a\over {m}}$, where $a$ is the s-wave scattering length and m is the mass of the atom. In the present work we considered these
three interactions are the same. Atom-molecular coupling strength via magnetic Feshbach resonance $\chi=\sqrt{4\pi\hbar^2a_{bg}(\Delta {\mu})(\Delta{B})/m}$,
where $a_{bg}$ is the background scattering length, $ (\Delta {\mu})$ is the difference between magnetic moments of two free atoms and the bound molecule and $(\Delta{B})$
is the width of the resonance. $\epsilon$ is the detuning from the resonance. The dipole-dipole interaction coefficients between two atoms and between two molecules are given as
$d_a=\mu_0\mu_d^2/{4\pi}$ and $d_m=\mu_0\mu_d'^2/{4\pi}$, $\mu_0$ is the permeability of free space, $\mu_d$ and $\mu_d'$
are the magnetic dipole moments of atoms and molecules respectively. The dipole-dipole interaction term $V ({\bf r}-{\bf r'})= {{1-3cos^2\theta}\over{({\bf r}-{\bf r'})^3}}$,
where $\theta$ is the angle between the vector $({\bf r}-{\bf r'})$ and the polarization direction taken along the axial symmetry axis z. 

Four coupled time-dependent GP like equations for dipolar atom-molecular system coupled via magnetic Feshbach resonance and trapped in a double well potential can be
obtained by varying mean field energy E with respect to $\psi_{al}^*$, $\psi_{ar}^*$, $\psi_{ml}^*$ and $\psi_{mr}^*$  and equating to the time derivative of the corresponding wave functions respectively  as follows:
\begin{eqnarray} i\hbar{\partial \psi_{al}\over \partial t}= {\delta E\over{\delta \psi_{al}^*}}; i\hbar{\partial \psi_{ar}\over \partial t}= {\delta E\over{\delta \psi_{ar}^*}};
i\hbar{\partial \psi_{ml}\over \partial t}= {\delta E\over{\delta
\psi_{ml}^*}}; i\hbar{\partial \psi_{mr}\over \partial t}= {\delta
E\over{\delta \psi_{mr}^*}}\end{eqnarray} Hence four
time-dependent coupled equations are written as:
$$i\hbar {\partial \psi_{al}({\bf r},t)\over \partial t}=[-{\hbar^2\nabla^2\over
2m}+v_a({\bf r})+\lambda_a |\psi_{al}({\bf r},t)|^2+\lambda_{am}|\psi_{ml}({\bf r},t)|^2 +$$
\begin {eqnarray}  d_a \int V ({\bf r}-{\bf r'}) |\psi_{al}({\bf r'},t)|^2 d{\bf r'}] \psi_{al}({\bf r},t) +
\chi \psi_{ml}({\bf r},t)\psi_{al}^*({\bf r},t) \end {eqnarray}

$$i\hbar {\partial \psi_{ar}({\bf r},t)\over \partial t}=[-{\hbar^2\nabla^2\over
2m}+v_a({\bf r})+\lambda_a |\psi_{ar}({\bf r},t)|^2+\lambda_{am}|\psi_{mr}({\bf r},t)|^2 +$$
\begin {eqnarray} d_a \int V ({\bf r}-{\bf r'}) |\psi_{ar}({\bf r'},t)|^2 d{\bf r'}] \psi_{ar}({\bf r},t) +
\chi \psi_{mr}({\bf r},t)\psi_{ar}^*({\bf r},t) \end {eqnarray}

$$i\hbar {\partial \psi_{ml}({\bf r},t)\over \partial t}=[-{\hbar^2\nabla^2\over
4m}+v_m({\bf r})+\lambda_m |\psi_{ml}({\bf r},t)|^2+\lambda_{am}|\psi_{al}({\bf r},t)|^2 +\epsilon+$$
\begin {eqnarray}  d_m \int V ({\bf r}-{\bf r'}) |\psi_{ml}({\bf r'},t)|^2 d{\bf r'}] \psi_{ml}({\bf r},t) +
{\chi\over 2} {\psi_{al}({\bf r},t)}^2\end {eqnarray}

$$i\hbar {\partial \psi_{mr}({\bf r},t)\over \partial t}=[-{\hbar^2\nabla^2\over
4m}+v_m({\bf r})+\lambda_m |\psi_{mr}({\bf r},t)|^2+\lambda_{am}|\psi_{ar}({\bf r},t)|^2 +\epsilon+$$
\begin {eqnarray} d_m \int V ({\bf r}-{\bf r'}) |\psi_{mr}({\bf r'},t)|^2 d{\bf r'}] \psi_{mr}({\bf r},t) +
{\chi\over 2} {\psi_{ar}({\bf r},t)}^2 \end {eqnarray}

Equations (3)-(6) can be written in dimensionless form \cite
{Gupta2009} by expressing energy, length, density and time in
units of oscillator energy $\hbar\omega_z$, oscillator length
$l_z=\sqrt{\hbar \over {m \omega_z}}$, ${l_z^{-3}}$ and
${\omega_z}^{-1}$ respectively:
$$i {\partial \psi_{al}({\bf r},t)\over \partial t}=[-{\nabla^2\over
2}+v'_a({\bf r})+\lambda'_a |\psi_{al}({\bf
r},t)|^2+\lambda'_{am}|\psi_{ml}({\bf r},t)|^2 +$$
\begin {eqnarray} d'_a U_{al}( {\bf r},t)] \psi_{al}({\bf r},t) +
\chi' \psi_{ml}({\bf r},t)\psi_{al}^*({\bf r},t)\end {eqnarray}

$$i {\partial \psi_{ar}({\bf r},t)\over \partial t}=[-{\nabla^2\over
2}+v'_a({\bf r})+\lambda'_a |\psi_{ar}({\bf
r},t)|^2+\lambda'_{am}|\psi_{mr}({\bf r},t)|^2 +$$
\begin {eqnarray} d'_a U_{ar}( {\bf r},t)] \psi_{ar}({\bf r},t) +
\chi' \psi_{mr}({\bf r},t)\psi_{ar}^*({\bf r},t) \end {eqnarray}

$$i {\partial \psi_{ml}({\bf r},t)\over \partial t}=[-{\nabla^2\over
4}+v'_m({\bf r})+\lambda'_m |\psi_{ml}({\bf
r},t)|^2+\lambda'_{am}|\psi_{al}({\bf r},t)|^2 + \epsilon'+$$
\begin {eqnarray}  d'_m U_{ml}({\bf r},t)] \psi_{ml}({\bf r},t) +
{\chi'\over 2} {\psi_{al}({\bf r},t)}^2\end {eqnarray}

$$i {\partial \psi_{mr}({\bf r},t)\over \partial t}=[-{\nabla^2\over
4}+v'_m({\bf r})+\lambda'_m |\psi_{mr}({\bf
r},t)|^2+\lambda'_{am}|\psi_{ar}({\bf r},t)|^2 +\epsilon'+$$
\begin {eqnarray}  d'_m U_{mr}( {\bf r},t)] \psi_{mr}({\bf r},t) +
{\chi'\over 2} {\psi_{ar}({\bf r},t)}^2\end {eqnarray}

Here 
\begin{equation}
\begin{aligned}
U_{al}({\bf r},t) =\int V ({\bf r}-{\bf r'}) |\psi_{al}({\bf r'},t)|^2 d{\bf r'}\\
U_{ar}({\bf r},t) =\int V ({\bf r}-{\bf r'}) |\psi_{ar}({\bf r'},t)|^2 d{\bf r'}
\end{aligned}
\end{equation}

\begin{equation}
\begin{aligned}
U_{ml}({\bf r},t) =\int V ({\bf r}-{\bf r'}) |\psi_{ml}({\bf
r'},t)|^2 d{\bf r'} \\U_{mr}({\bf r},t) =\int V ({\bf r}-{\bf r'})
|\psi_{mr}({\bf r'},t)|^2 d{\bf r'}
\end{aligned}
\end{equation}

$$v'_a({\bf r})= {1\over 2}({\rho^2\over \lambda^2}+z^2) +Ae^{-\kappa z^2}; v'_m({\bf r})= ({\rho^2\over \lambda^2}+z^2) +Ae^{-\kappa z^2};
d'_a={d_a\over{l_z^2 \hbar \omega_z}};$$ $$ d'_m={d_m\over{l_z^2 \hbar \omega_z}};\lambda'_a=
{\lambda_a\over {l_z^2 \hbar \omega_z}};
\lambda'_a=\lambda'_m=\lambda'_{am};\chi'={\chi\over{l_z \hbar
\omega_z}};\epsilon'={\epsilon\over{ \hbar \omega_z}}$$

\subsection{\bf{Derivation of one dimensional coupled equations
to study the dynamics of dipolar atom-molecular BECs in double
well}:}

To derive the coupled equations in the axial direction (z) from
the 3D equations (7)-(10), we consider strong confinement in the
radial direction i.e $\omega_z<<\omega_\rho$ and hence the
asymmetry parameter $\lambda={\omega_z\over \omega_\rho}<<1$. Due
to the strong confinement it is assumed that the BECs in the
radial trap are confined in the ground state and the corresponding
wavefunctions are given as:
\begin{eqnarray}
{\psi_{a0}(\rho)}={1\over{\sqrt {\pi\lambda}}}
e^{-{\rho^2\over{2\lambda}}}   ;\psi_{m0}(\rho)={\sqrt{2\over
{\pi\lambda}}} e^{-{\rho^2\over{\lambda}}}
\end{eqnarray}
respectively. Hence the total wavefunction for atoms and molecules
in the left and right wells are given as:
\begin{eqnarray}
{\psi_{al}({\bf r},t)=\psi_{a0}(\rho)}\psi_{al}(z,t);
\psi_{ar}({\bf r},t)=\psi_{a0}(\rho)\psi_{ar}(z,t)
\end{eqnarray}

\begin{eqnarray}
{\psi_{ml}({\bf r},t)=\psi_{m0}(\rho)}\psi_{ml}(z,t);
\psi_{mr}({\bf r},t)=\psi_{m0}(\rho)\psi_{mr}(z,t)
\end{eqnarray}
Hence by substituting equations (14) and (15) in equations
(7)-(10), multiplying atomic equations by $\psi_{a0}(\rho)$ and
molecular equations by $\psi_{m0}(\rho)$ on the left respectively
and integrating over $\rho$ one dimensional coupled equations for
atoms and molecules can be obtained as:

$$i {\partial \psi_{al}(z,t)\over \partial t}=[-{1\over 2}{\partial^2\over{\partial z^2}}
+v_a(z)+\lambda''_a
|\psi_{al}(z,t)|^2+\lambda''_{am}|\psi_{ml}(z,t)|^2 +$$
\begin{eqnarray}
d'_a U_{al}(z,t)] \psi_{al}(z,t) + \chi''
\psi_{ml}(z,t)\psi_{al}^*(z,t)
\end{eqnarray}

$$i {\partial \psi_{ar}(z,t)\over \partial t}=[-{1\over 2}{\partial^2\over{\partial z^2}}
+v_a(z)+\lambda''_a
|\psi_{ar}(z,t)|^2+\lambda''_{am}|\psi_{mr}(z,t)|^2 +$$
\begin{eqnarray}  d'_a U_{ar}(z,t)] \psi_{ar}(z,t) +
\chi'' \psi_{mr}(z,t)\psi_{ar}^*(z,t)\end{eqnarray}

$$i {\partial \psi_{ml}(z,t)\over \partial t}=[-{1\over 4}{\partial^2\over{\partial z^2}}
+v_m(z)+\lambda''_m
|\psi_{ml}(z,t)|^2+\lambda''_{am}|\psi_{al}(z,t)|^2 +\epsilon'+$$
\begin{eqnarray} d'_m U_{ml}(z,t)] \psi_{ml}(z,t) +
\chi'' \psi_{al}^2(z,t)\end{eqnarray}

$$i {\partial \psi_{mr}(z,t)\over \partial t}=[-{1\over 4}{\partial^2\over{\partial z^2}}
+v_m(z)+\lambda''_m
|\psi_{mr}(z,t)|^2+\lambda''_{am}|\psi_{ar}(z,t)|^2 +\epsilon'$$
\begin{eqnarray} d'_m U_{mr}(z,t)] \psi_{mr}(z,t) +
\chi'' \psi_{ar}^2(z,t) \end{eqnarray} where
$v_a(z)={1\over2}z^2+Ae^{-\kappa z^2};$ $v_m(z)=z^2+Ae^{-\kappa
z^2};$ $\lambda''_a={\lambda'_a\over{2\pi\lambda}};$
$\lambda''_a=\lambda''_m=\lambda''_{am};$
$\chi''={\chi'\over{\sqrt{2\pi\lambda}}}$ ;

\begin{equation}
\begin{aligned}
U_{al}(z.t)=\int V (|z-z'|) |\psi_{al}(z',t)|^2 dz'\\
U_{ar}(z,t) =\int V (|z-z'|) |\psi_{ar}(z',t)|^2 dz'
\end{aligned}
\end{equation}
\begin{equation}
\begin{aligned}
U_{ml}(z,t) =\int V (|z-z'|) |\psi_{ml}(z',t)|^2 dz'\\
U_{mr}(z,t)=\int V (|z-z'|) |\psi_{mr}(z',t)|^2 dz'
\end{aligned}
\end{equation}

\subsection
{\bf{Derivation of dipole-dipole interaction potential in one 
dimension(z)}}
The dipole-dipole interaction potential in axial direction can be
obtained by taking the Fourier transforms of 3D potentials given
in equations (11) and (12) as follows: Fourier transform of
dipole-dipole interaction is given as \cite {Ronen2006}
\begin{eqnarray}
F[V({\bf r}-{\bf r'})]={4\pi \over 3}
{[{3k_z^2\over{k_{\rho}^2+k_z^2}}-1]}
\end{eqnarray}
Fourier transforms of densities are given
as:\begin{eqnarray}n_{a0}({\bf k_\rho})=\int_0^{\infty} e^{i {\bf
{k_\rho}} \cdot {\bf {\rho}}}{n_{a0}(\rho)} d{\bf
\rho}=e^{-{{k^2_\rho}\lambda\over 4}}\end{eqnarray}
\begin{eqnarray} 
n_{m0}({\bf{k_\rho}})=\int_0^{\infty} e^{i{\bf k_\rho \cdot \rho}}
{n_{m0}(\rho)}{d{\bf \rho}}=e^{-{{k^2_\rho}\lambda\over 8}}
\end{eqnarray}
\begin{eqnarray}
n_{ai}(k_z,t)={\int_{-\infty}^\infty}{dz e^{ik_zz} n_{ai}(z,t)}
\end{eqnarray}
\begin{eqnarray}
n_{mi}(k_z,t)={\int_{-\infty}^\infty}{dz e^{ik_zz} n_{mi}(z,t)}
\end{eqnarray}
Here the densities in coordinate space are given as
$n_{j0}(\rho)=|\psi_{j0}(\rho)|^2$ ;
$n_{ai}(z,t)=|\psi_{ai}(z,t)|^2$ and
$n_{mi}(z,t)=|\psi_{mi}(z,t)|^2$ where the suffix 'i' is for
suffix 'l' and 'r' corresponding to left and right well and  the
suffix 'j' corresponds to atoms and molecules. Hence using
equations (22)-(26) the dipole-dipole interaction potentials in
momentum space as a function of $({\bf{k_\rho}},k_z,t)$ is
obtained. Then performing the integration over $\bf{k_\rho}$ and
taking the inverse Fourier transform over $k_z$ one can get the
interactions in one dimension (z) as follows:

\begin{eqnarray}
U_{ai}(z,t) =\int_{-\infty}^{\infty} 
{dk_z\over 2\pi}e^{-ik_zz}U_{ai}(k_z)n_{ai}(k_z,t)
\end{eqnarray}
\begin{eqnarray}
U_{mi}(z,t) =\int_{-\infty}^{\infty} 
{dk_z\over 2\pi}e^{-ik_zz}U_{mi}(k_z)n_{mi}(k_z,t)
\end{eqnarray} where

\begin{equation} 
\begin{split}
U_{ai}(k_z) & = {4\pi\over3}{1\over(2\pi)^2}\int_0^\infty {dk_\rho}{[{3k_z^2\over{k_{\rho}^2+k_z^2}}-1]}{|n_{a0}(\bf{k_\rho})|^2} \\
 & = {4\pi\over3}{1\over(2\pi)^2}\int_0^\infty {dk_\rho}
{[{3k_z^2\over{k_{\rho}^2+k_z^2}}-1]}e^{-{k^2_\rho \lambda\over
2}}\\
 & = {4\pi\over3}{1\over{2\pi\lambda}}\int_0^{\infty} dv
[{3\xi^2\over{v+\xi^2}}-1]e^{-v}\\
& = {4\pi\over3}{1\over{2\pi\lambda}} S_a(\xi)
\end{split}
\end{equation}
where $S_a(\xi)=[{3\xi^2\over{v+\xi^2}}-1]e^{-v};$
$\xi=k_z\sqrt{\lambda\over 2}$ and $v=k_\rho^2{\lambda\over 2}$

\begin{equation} 
\begin{split}
U_{mi}(k_z) & ={4\pi\over3}{1\over(2\pi)^2}\int_0^\infty {dk_\rho}{[{3k_z^2\over{k_{\rho}^2+k_z^2}}-1]}{|n_{m0}\bf{k_\rho}|^2} \\
 & = {4\pi\over3}{1\over(2\pi)^2}\int_0^\infty {dk_\rho}{[{3k_z^2\over{k_{\rho}^2+k_z^2}}-1]}e^{-{k^2_\rho \lambda\over 4}}\\
 & ={4\pi\over3}{2\over{2\pi\lambda}}\int_0^{\infty} du [{3\zeta^2\over{u+\zeta^2}}-1]e^{-u}\\
& ={4\pi\over3}{2\over{2\pi\lambda}} S_m(\zeta)
\end{split}
\end{equation}
 where
$S_m(\zeta)= [{3\zeta^2\over{u+\zeta^2}}-1]e^{-u};$ $\zeta=k_z{\sqrt{\lambda}\over 2}$ and $u=k_\rho^2{\lambda\over 4}$\\
Therefore by equating equations (20) and (27) and substituting
equation (29) one can write down the dipole-dipole interaction
term for atoms in left and right wells in equations (16) and (17)
as:
\begin{equation}
\begin{split}
\int V (|z-z'|) |\psi_{ai}(z',t)|^2 dz' & =\int_{-\infty}^{\infty} {dk_z\over 2\pi}e^{-ik_zz}U_{ai}(k_z)n_{ai}(k_z,t)\\
& = {4\pi\over3}{1\over{2\pi\lambda}}\int_{-\infty}^{\infty} {d k_z\over 2\pi}e^{-ik_zz}S_a(\xi)n_{ai}(k_z,t)
\end{split}
\end{equation}

 Similarly by equating equations (21) and (28) and substituting equation (30) the dipole-dipole interaction term for molecules in left and right wells in equations (18) and (19) can be given as:
 \begin{equation}
\begin{split}
 \int V (|z-z'|) |\psi_{mi}(z',t)|^2 dz' & =\int_{-\infty}^{\infty} {dk_z\over 2\pi}e^{-ik_zz}U_{mi}(k_z)n_{mi}(k_z,t)\\
& = ={4\pi\over3}{2\over{2\pi\lambda}}\int_{-\infty}^{\infty} {d
k_z\over 2\pi}e^{-ik_zz}S_m(\zeta)n_{mi}(k_z,t)
\end{split}
\end{equation}

The normalization condition used is $ N=N_a(t)+2N_m(t)$ where N is
the initial total number of atoms, $N_a(t)$ and $N_m(t)$ are the
total number of atoms and molecules respectively at each instant
of time; $N_a(t)=N_{al}(t)+N_{ar}(t)$;
$N_m(t)=N_{ml}(t)+N_{mr}(t)$. $N_{al}(t)$ and $N_{ar}(t)$ are the
number of atoms in left and right well respectively at time t and
are defined as: $N_{al}(t)=\int_{-\infty}^0 dz n_{al}(z,t)$; and
$N_{ar}(t)=\int_0^{\infty}dz n_{ar}(z,t)$. Similarly the number of
molecules in left and right well at time t are given as:
$N_{ml}(t)=\int_{-\infty}^0 dz n_{ml}(z,t)$; and
$N_{mr}(t)=\int_0^{\infty} dz n_{mr}(z,t)$ respectively.
Population imbalance is defined as
$z_a(t)=(N_{al}(t)-N_{ar}(t))/N_a$ for atoms and for molecules
$z_m(t)=(N_{ml}(t)-N_{mr}(t))/N_m$ where $N_a$ and $N_m$ are the
initial total number of atoms and molecules respectively and
$N=N_a+2N_m$.

\section {Numerical Calculation}
To study the dynamical behaviour of coupled atom-molecular BEC
system with long range dipole-dipole interaction and short range
contact interactions between particles we solved four one
dimensional (in axial z direction) coupled equations (16)-(19)
numerically using Crank-Nicholson scheme to discretize the
equations by using a space step $h=.005$ and time step $\delta
t=5\times {10^{-8}}$. For initial input wavefunctions we have
solved the time-independent atom-molecular coupled (two) equations
by imaginary time method without considering the dipole-dipole
interaction. To obtain initial guess wavefunctions in left and
right wells we have divided the total wavefunctions for atom and
molecules $\psi_{ag}(z)$ and $\psi_{mg}(z)$ respectively in the
range $-\infty < z < \infty$ into two parts. The initial guess
wavefuncions for atoms in left and right wells are defined as:
$\psi_{al}(z)$=$\psi_{ag}(z)$ for the values of z from $-\infty$
to 0 and $\psi_{ar}(z)$=$\psi_{ag}(z)$ for the values of z from 0
to $\infty$ respectively. Similarly initial guess wavefunctions
for molecules in left and right wells are defined as:
$\psi_{ml}(z)$=$\psi_{mg}(z)$ for the values of z from $-\infty$
to 0 and $\psi_{mr}(z)$ =$\psi_{mg}(z)$ for the values of z from 0
to $\infty$ respectively.

Initial total number of atoms is taken to be 5000 which is
distributed unequally in left and right wells. The number of atoms
in left and right wells is taken to be 1500 and 1000 respectively.
Initial number of molecules has been chosen to be half of the
number of atoms. Hence the number of molecules in the left and
right wells are 750  and 500 respectively. Hence $N_{al}(0)=1500$,
$N_{ar}(0)=1000$, $N_{ml}(0)=750$, $N_{mr}(0)=500$ and
$z_a(0)=z_m(0)=0.2$ for this calculation. To ensure specific
initial number of atoms and molecules in left and right wells we
have normalized $\psi_{al}(z)$, $\psi_{ar}(z)$, $\psi_{ml}(z)$ and
$\psi_{mr}(z)$ accordingly.  This type of unequal population
distribution can be reached by considering the asymmetric double
well potential \cite {Adhikari2014,Albiez2005, Gati2007}. The
relative phase between atomic and molecular condensates is
considered to be $\pi$ in both the wells. However the atomic
condensates in the left and right well are chosen to be in phase
and similarly the relative phase is zero for molecular
condensates.  The convergence of the wavefunctions in this work
has been checked to be $\leq{10^{-8}}$.

For simplicity we have assumed atom-atom, atom-molecule and
molecule-molecule interactions are the same i.e.
$\lambda''_a=\lambda''_m=\lambda''_{am};$ and for this bosonic
system the short range contact interaction is repulsive in nature.
After simplification the coefficient for atom-atom contact
interaction in equations (16) and (17) can be reduced to
$\lambda''_a={2a\over \lambda}$. In the present work we have
chosen that the dipole-dipole interactions between molecules in
left and right wells are the same as those of atoms i.e. $d_m'
U_{mi}(k_z)=d_a' U_{ai}(k_z)$ where $U_{ai}(k_z)$ and
$U_{mi}(k_z)$ are given in equations (29) and (30) respectively.
Here we have considered that the dipoles are aligned along the
polarization axis \cite {Xiong2009}, such that the dipole-dipole
interaction is attractive in nature. The calculation of
dipole-dipole interaction potential for the atoms and molecules
given in equations (27) and (28) respectively has been done in two
steps:(i) we have considered the functions $S_a(\xi)$ and
$S_m(\zeta)$ are equal and $S_a(\xi)$ has been expressed as
Exponential functions which has been multiplied by the Fourier
transforms of atomic density $n_{ai}(k_z,t)$ and molecular density
$n_{mi}(k_z,t)$ for atoms and molecules respectively. Then (ii)
the  inverse Fourier transform of the products has been done to
obtain $U_{ai}(z,t)$ and $U_{mi}(z,t)$ as given in equations (27)
and (28). The Fourier transforms of densities and the inverse
Fourier transforms have been done numerically. If a dipolar length
scale is defined as $a_{dd}={{\mu_0 \mu_d^2}m\over{12\pi\hbar^2}}$
\cite {Adhikari2014} then the coefficient of dipole-dipole
interaction for atoms become ${d_a'}{4\pi\over 3}{1\over {2\pi
\lambda}}={2a_{dd}\over \lambda}$.

In this model calculation we have chosen different parameters as:
(i) the coefficient of dipole-dipole interaction=0.75 x atom-atom
contact interaction, (ii) the s-wave scattering length of atoms
$'a'$ has been chosen to be $a=20a_0$ and the mass of atoms m is
taken as that of Cr atoms, (iii) the angular frequency
$\omega_z=2\pi\times 194 Hz$ and the asymmetry parameter
$\lambda=0.11$, (iv) the double well parameters $A=15$ and
$\kappa=10$ respectively, (v) the  value of the atom-molecular
coupling strength in dimensionless form is taken as
$\chi''=0.23012$ \cite{Gupta2009} and
 (vi) the detuning $\epsilon=8\times 10^4 Hz$.

 \section {Results and Discussions}
To study the dynamical behaviour of atomic and molecular coupled
BECs in a double well potential in presence of dipole-dipole long
range interaction as well as short range contact interaction four
time-dependent coupled GP like equations have been solved by
applying the steepest descent method in the Crankâ€“Nicholson
discretization scheme. Details of Crank-Nicholson scheme has been
discussed previously \cite {Gupta2009,Sreoshi2021}. The effect of
dipole-dipole long range interaction and atom-molecular coherence
through Feshbach resonance on the transmit/oscillation/trapping of
atomic and molecular population between two wells has been
studied. The total number of atoms and molecules in left well
$(N_{al}(t), N_{ml}(t))$ and right well $(N_{ar}(t) , N_{mr}(t))$
has been calculated at each instant of time in presence and
absence of dipole-dipole interaction as well as in presence and
absence of coherent coupling between atoms and molecules and
plotted as a function of time. Corresponding population imbalance
for atoms $z_a(t)$ and molecules $z_m(t)$ as a function of time
has also been presented. To demonstrate these effects on the
density profile of atoms, atomic density as a function of z has
been plotted at times in the range of 76 to 120. The value of t
mentioned here is in units of $\omega_z^{-1}$.

\subsection{Effect of coherent coupling on the dynamics of population in the presence of dipole-dipole interaction:}
Evolution of total number of atoms and molecules in left and right
wells is shown in Fig.1. It shows the dynamical behaviour of atoms
and molecules in two wells considering the long range
dipole-dipole interaction in presence and absence of coherent
coupling $(\chi)$ between atoms and molecules. In these figures
(Fig.1(a) and Fig.1(b)) upper and lower panel show dynamics of
atoms and molecules respectively. Both in Fig.1(a) and Fig.1(b)
atomic and molecular populations are found to oscillate initially.
But in Fig.1(a) (in presence of coherent coupling $\chi$)
transmission of atoms from left to right well starts after time $t
\approx {55}$, when the population in right well becomes much
higher than that in the left well and this transmission persists
up to the time $t \approx {125}$ with some oscillations where
populations become equal or very close to each other. However this
transient transmission effect is washed out in Fig.1(b) (in
absence of $\chi$) where atoms are self trapped in left well for
the period of time t=75 to 175 approximately with some
oscillations where populations become equal or close to each
other. These features are also demonstrated in Fig.2(a) and
Fig.2(b) where the population imbalance $z_a(t)$ of atoms
corresponding to Figs.1(a) and 1(b) has been plotted. Fig.2(a)
shows partial transmission for a time duration of t=55 to 125
approximately when most of the time $z_a(t)$ remains negative
except at some points where it touches zero. Whereas $z_a(t)$ in
Fig.2(b) is positive most of the time in the period t=75 to 175
with some oscillations touching zero. Therefore the presence of
coherent coupling between atoms and molecules with dipole-dipole
long range interaction prefers transient transmission of atoms
dominantly to the right well in contrary to transient self
trapping in left well in absence of it. This type of oscillation
with time in the atomic population imbalance for non-dipolar
atom-molecular coupled BECs has been obtained in case of initial
unequal population distribution of atoms and molecules in two
wells and for moderately strong tunnelling effect  \cite
{Atsushi2009}.

\begin{figure}
\begin{subfigure}[t]{0.45\textwidth}
\centering
\includegraphics[width=\textwidth]{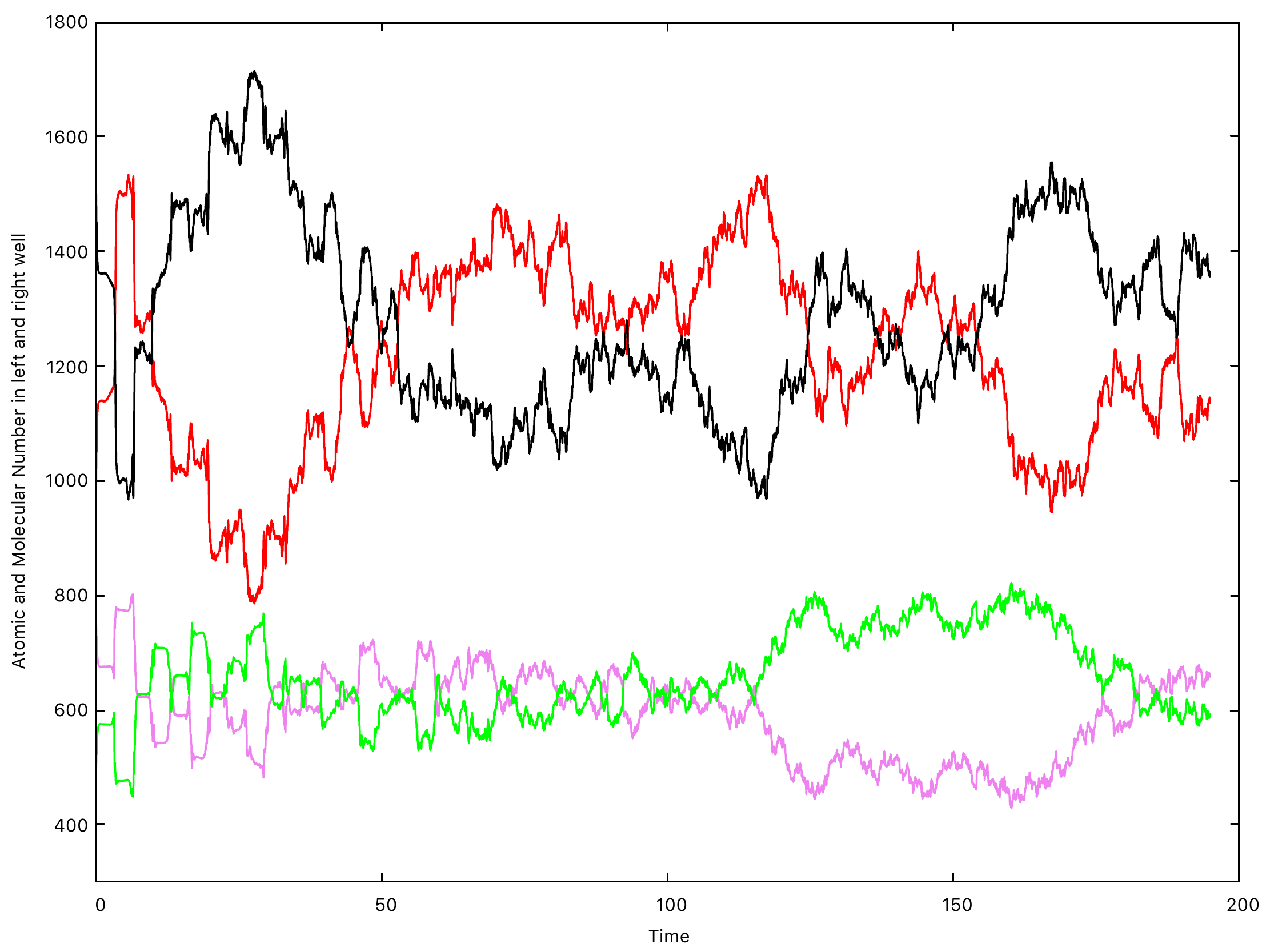}
{\caption*{(a)}}
\end{subfigure}
\hfill
\begin{subfigure}[t]{0.45\textwidth}
\includegraphics[width=\textwidth]{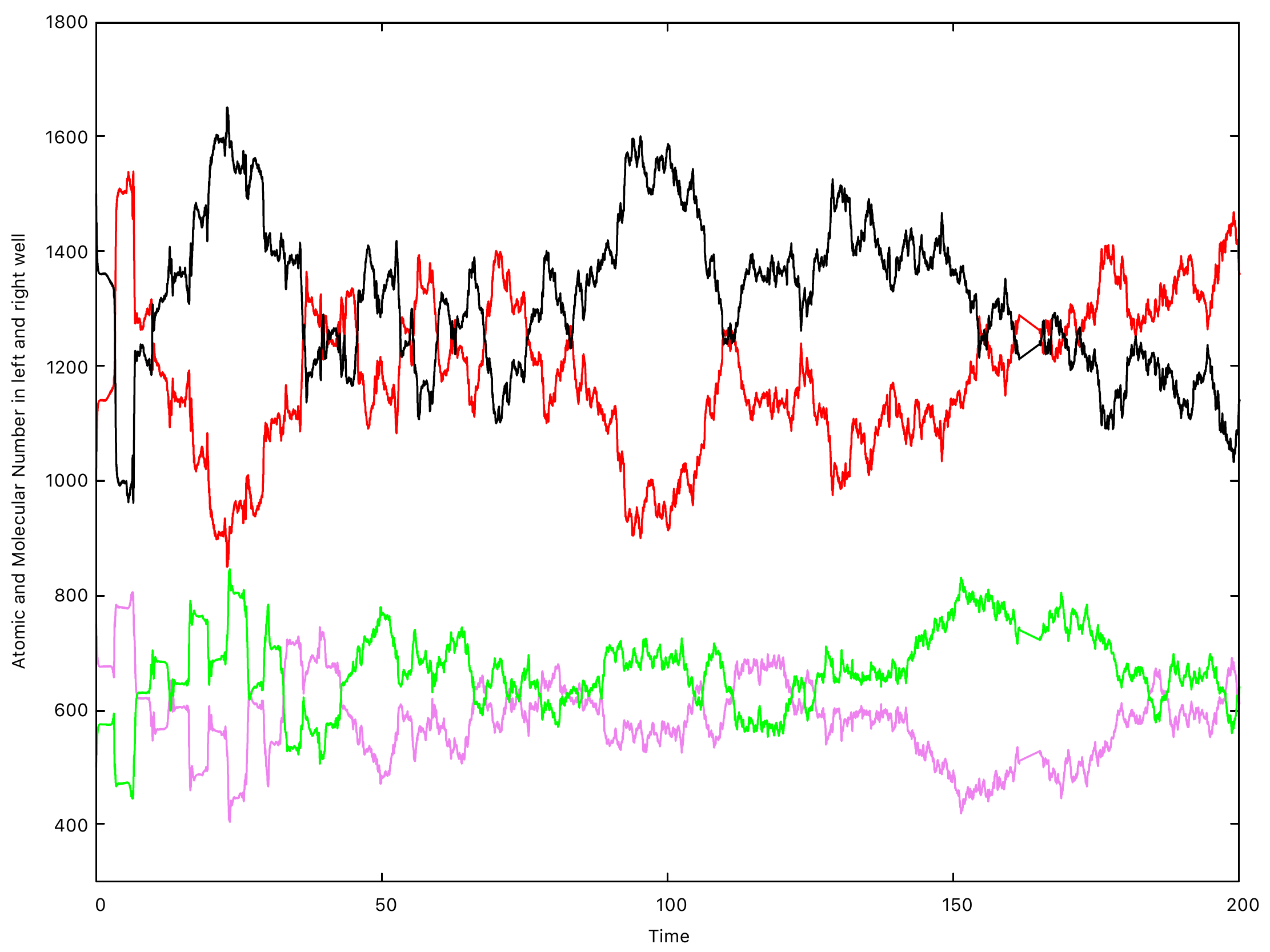}
{\caption*{(b)}}
\end{subfigure}

{\caption*{Figure 1: Atomic (upper panel) and Molecular (lower panel) population distribution as a function of time.
Total number of atoms and molecules in left (black, violet) and
right (red, green) wells are given for (a) $\chi \neq 0$ and
$a_{dd}=.75a$ and (b) $\chi=0$ and $a_{dd}=.75a$ respectively. In
all the figures 'Time' is given in units of $\omega_z^{-1}$.}}
\end{figure}

Dynamical behaviour of molecules (lower panel) in Fig.1(a) shows
initial oscillations and the transfer of population to right well
starts at t=110 approximately and remains higher than that in the
left well up to the time t=185 approximately. The feature of
transient transmission of molecular population exists in presence
of coherent coupling as in the case of atomic transmission, but it
takes double the time that for atoms to start. In absence of
coherent coupling transfer of molecular population (lower panel of
Fig.1(b)) occurs at larger time t=130 and the prominence of
transfer persists for a shorter time than that in case of Fig.1(a)
(lower panel).
\begin{figure}
\begin{subfigure}[t]{0.45\textwidth}
\centering
\includegraphics[width=\textwidth]{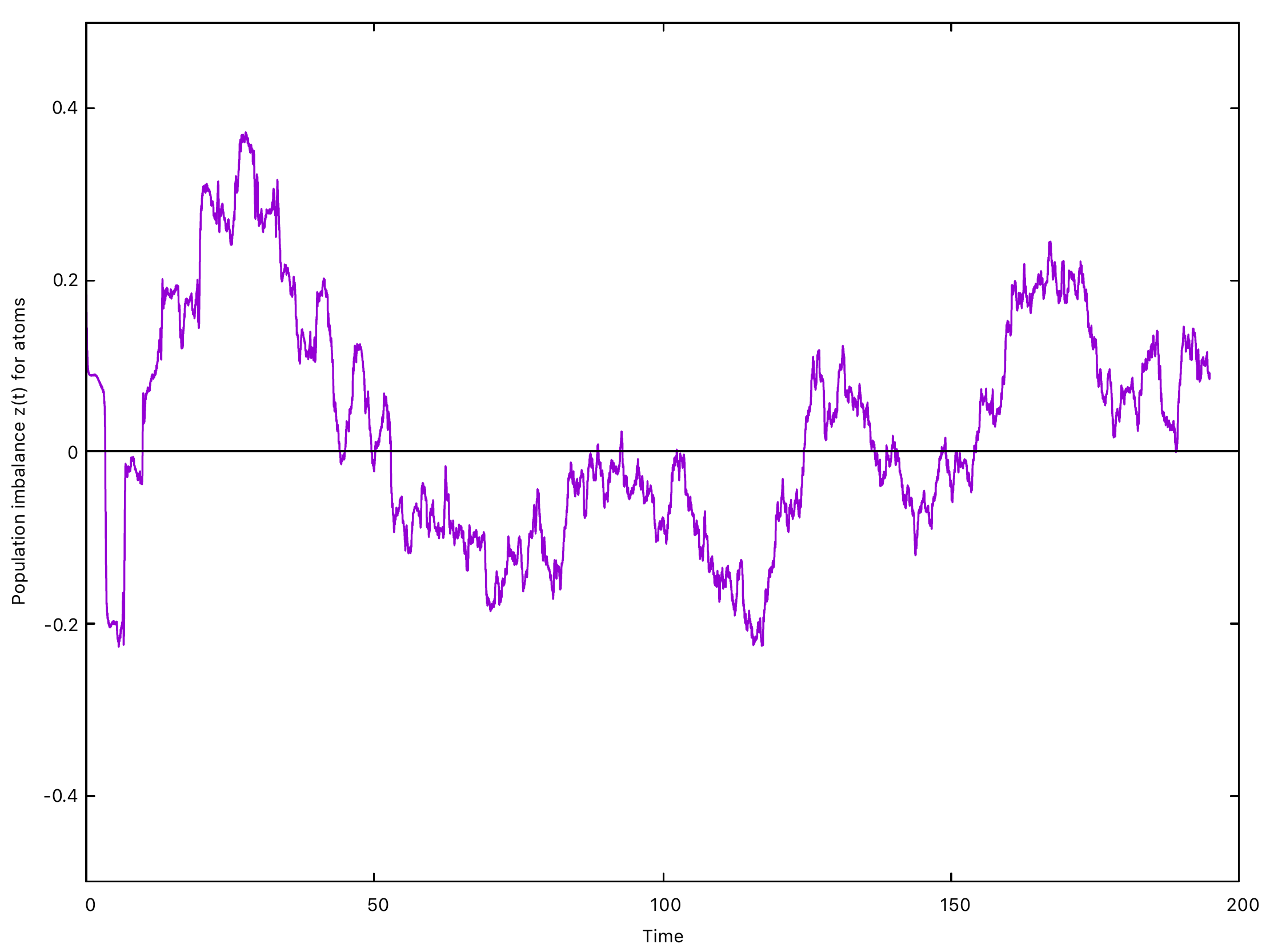}
{\caption*{(a)}}
\end{subfigure}
\hfill
\begin{subfigure}[t]{0.45\textwidth}
\includegraphics[width=\textwidth]{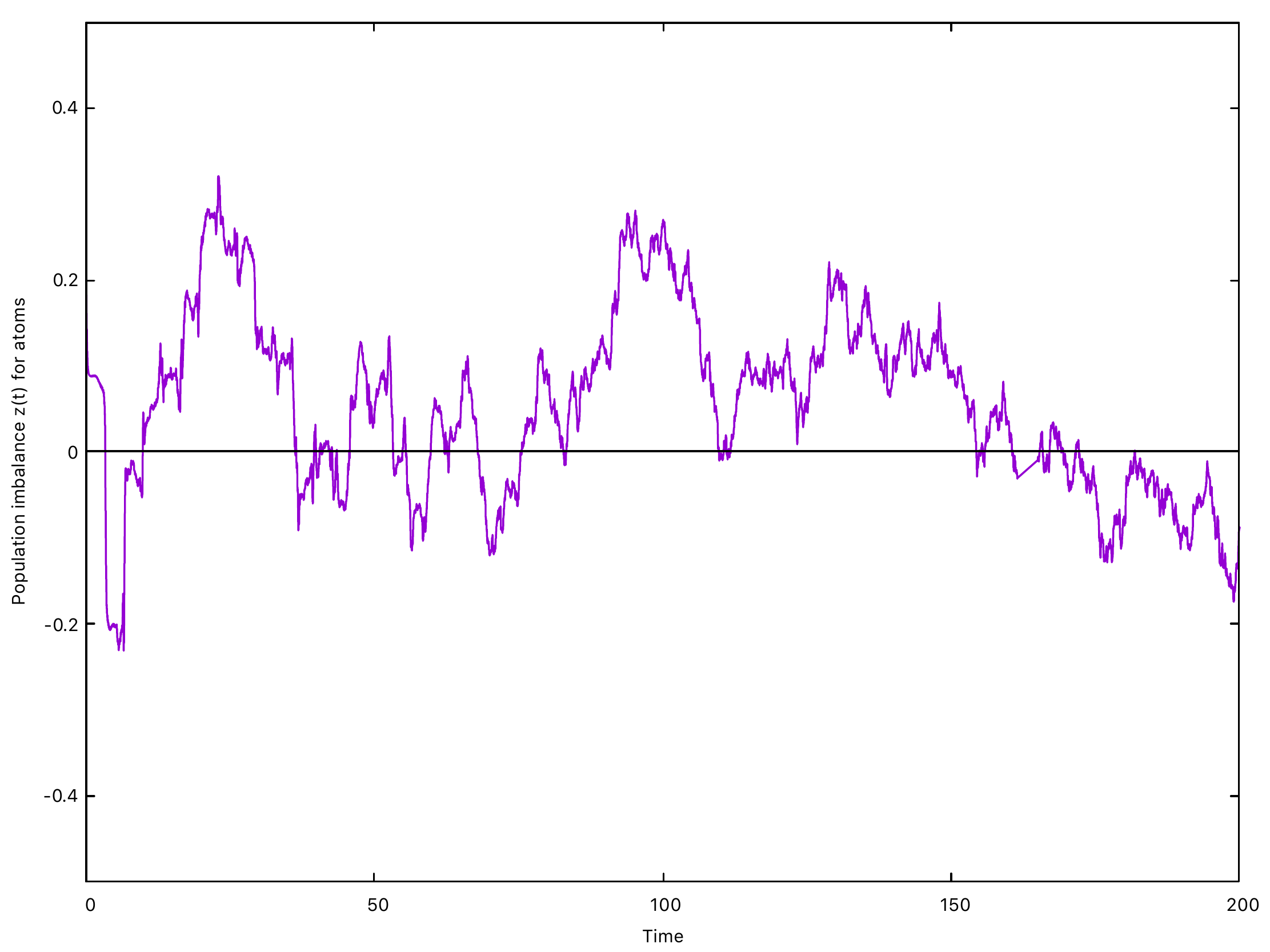}
{\caption*{(b)}}
\end{subfigure}

{\caption*{Figure 2: Population imbalance for atoms for (a) $\chi\neq 0$ and $ a_{dd}=.75a$ and (b) for $\chi=0$ and $ a_{dd}=.75a$}}
\end{figure}

Molecular tunneling effect in non-dipolar atomic-molecular coupled
(via photo association) BECs has been discussed \cite {Jing2006}
previously choosing certain parameters. It has been shown that the
molecular number in the right well dominates over that in the left
well and oscillates with time (similar to the results shown here)
for small atom-molecular coupling.

\begin{figure}
\begin{subfigure}[t]{0.45\textwidth}
\centering
\includegraphics[width=\textwidth]{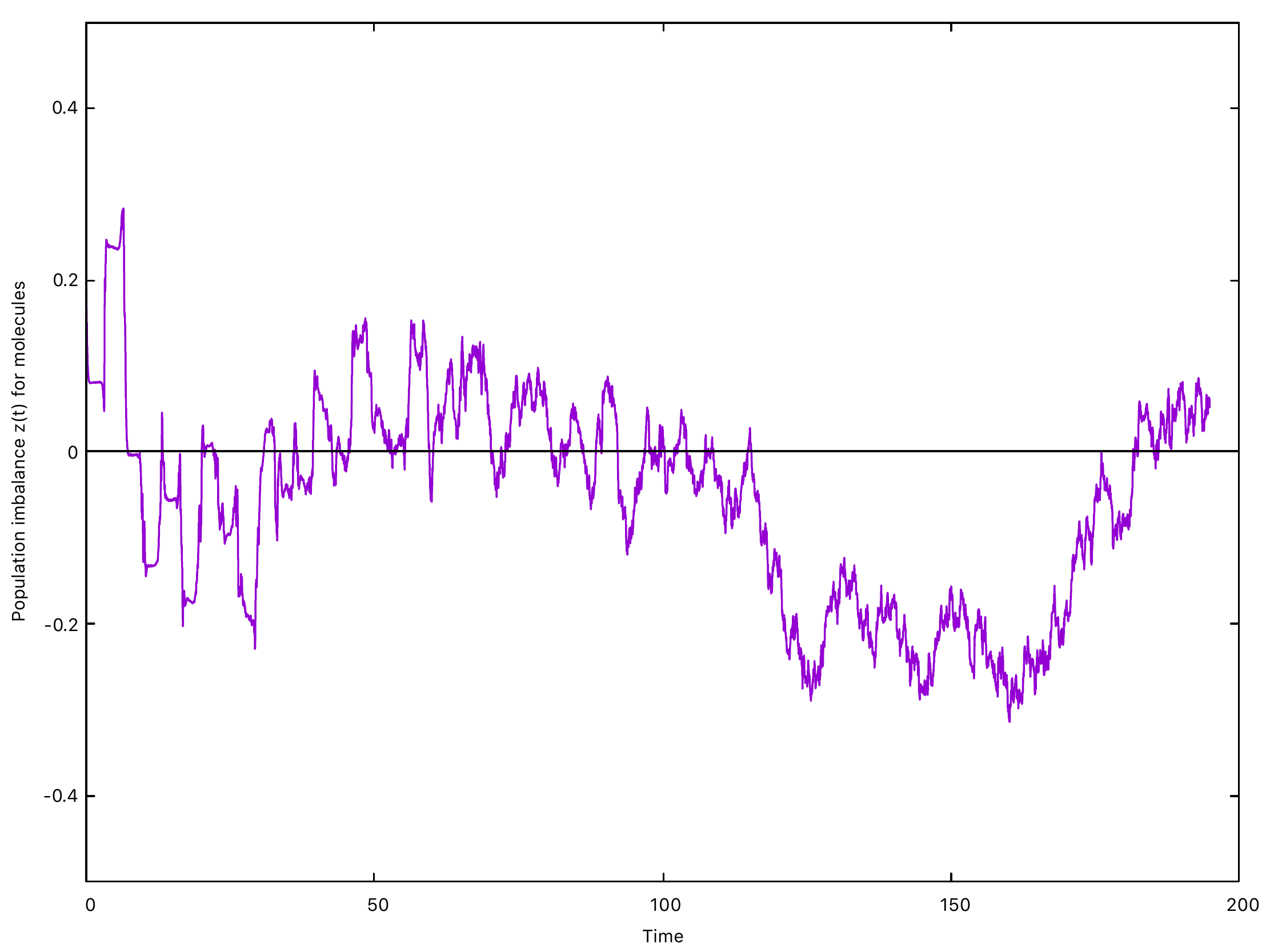}
{\caption*{(a)}}
\end{subfigure}
\hfill
\begin{subfigure}[t]{0.45\textwidth}
\includegraphics[width=\textwidth]{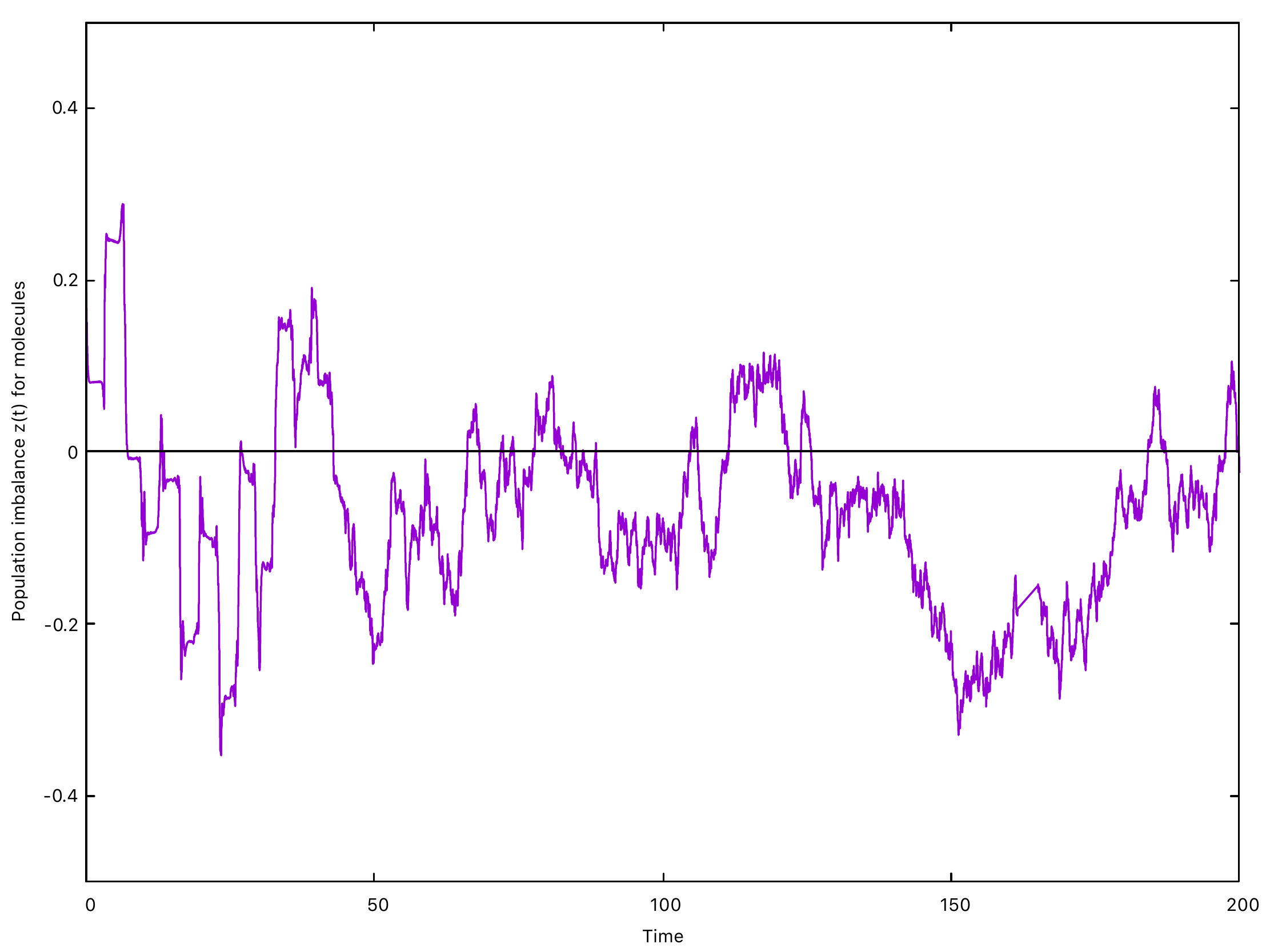}
{\caption*{(b)}}
\end{subfigure}

{\caption*{Figure 3: Population imbalance for molecules for (a) $\chi\neq 0$ 
and $ a_{dd}=.75a$ and (b) $\chi=0$ and $ a_{dd}=.75a$}}
\end{figure}

The transmission effect on molecules as shown in figures 1(a) and
1(b) has been supported by the plot of population imbalance of
molecules $z_m(t)$ in Fig.3(a) and Fig.3(b) respectively.
Therefore as in the case of atoms coherent coupling between atoms
and molecules leads to transient transmission of molecules in the
right well and the population in right well becomes larger than
that in the left well for a period of time mentioned above.
However this transient transmission time is shortened in absence
of coherent coupling.

Therefore from these results it is found that transmission of both
the atoms and molecules for a period of time is facilitated in
presence of atom molecular coherence in a dipolar atom-molecular
coupled BEC system. Whereas in absence of it this transient
transmission effect is damped.

\subsection{Effect of coherent coupling on the dynamics of population in the absence of dipole-dipole interaction:}
To investigate the effect of coherent coupling on the dynamics of atoms and molecules in absence of dipole-dipole interaction we repeated the calculation with $a_{dd}=0$ and plotted the total number of atoms and molecules (upper and lower panel respectively) as a function of time in Fig.4(a) and 4(b) in presence and absence of coherent coupling ($\chi$) respectively. It is found that for $\chi=0$ and $a_{dd}=0$, self trapping of atomic population (Fig.4(b)) is present for a long period of time t=30 to 140 approximately except at $t\approx{118}$ where a small population exchange occurs. This feature of self trapping is also evident in the plot of corresponding population imbalance $z_a(t)$ in Fig.5(b). The self-trapping effect has been obtained previously in our group \cite {Saha2019} for non-dipolar cold atoms trapped in a double well in case of strong confinement i.e. the asymmetry parameter $\lambda<<1$. 
\begin{figure}
\begin{subfigure}[t]{0.45\textwidth}
\centering
\includegraphics[width=\textwidth]{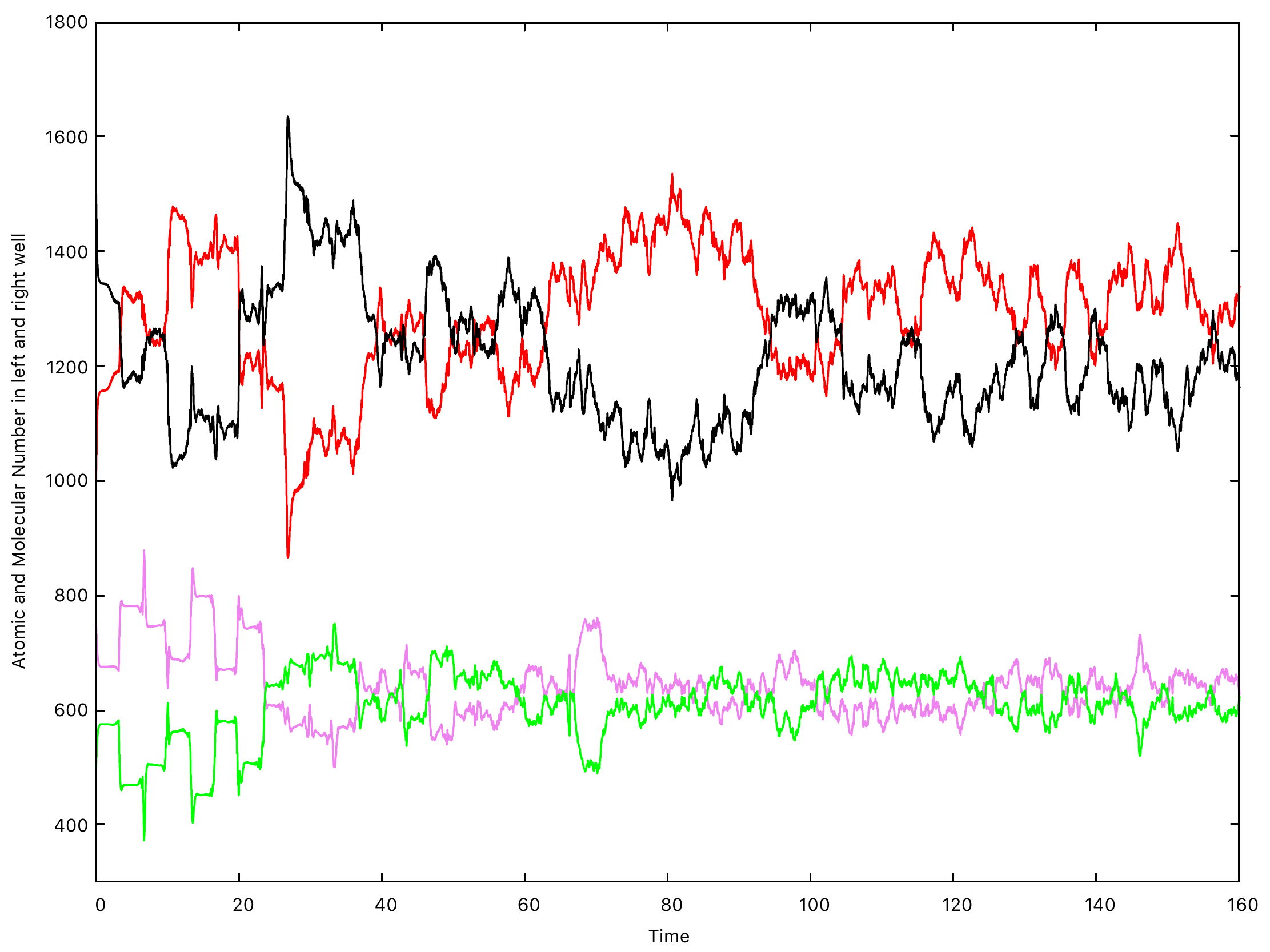}
{\caption*{(a)}}
\end{subfigure}
\hfill
\begin{subfigure}[t]{0.45\textwidth}
\includegraphics[width=\textwidth]{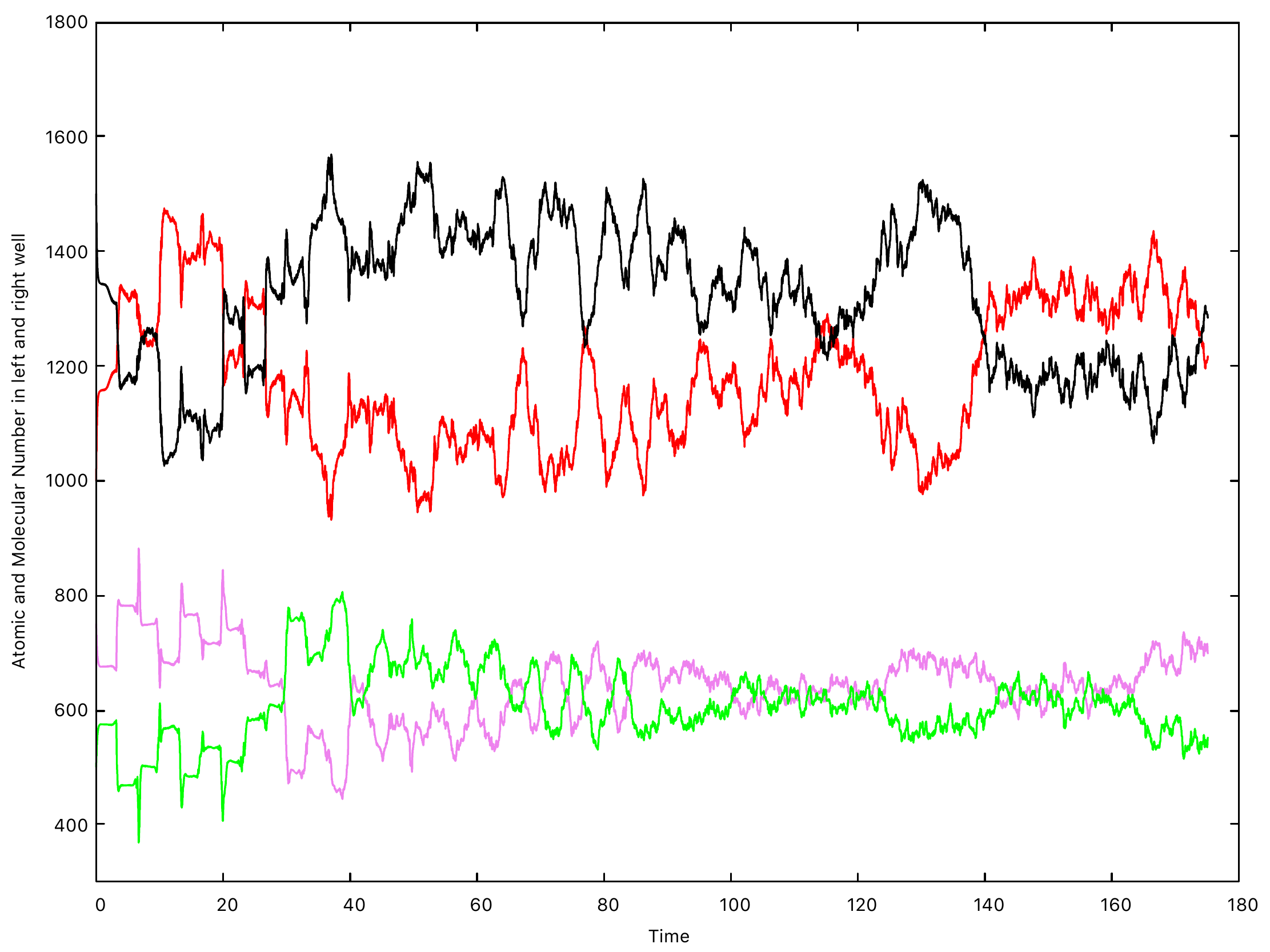}
{\caption*{(b)}}
\end{subfigure}

{\caption*{Figure 4: Atomic (upper panel) and Molecular (lower panel) population distribution as a function of time.
Total number of atoms and molecules in left (black, violet) and
right (red, green) wells are given for (a) $\chi \neq 0$ and
$a_{dd}=0$ and (b) $\chi=0$ and $a_{dd}=0$ respectively.}}
\end{figure}

It is shown here that in presence of coherent coupling this self
trapping is disturbed and transmission of atomic population is
prominent for a shorter period of time  (65 to 95) in Fig.4(a)
than that in Fig.1(a). However a small transfer of population is
present during the time 115 to 150 with small oscillations. This
can be seen in the plot of corresponding population imbalance
$z_a(t)$ in Fig.5(a). Therefore from these figures it is found
that the  transient self trapping in the left well and  the
transient transmission of population to be trapped in the right
well can be induced in absence and presence of coherent coupling
respectively, even when the long range dipole-dipole interaction
is considered to be zero.
\begin{figure}
\begin{subfigure}[t]{0.45\textwidth}
\centering
\includegraphics[width=\textwidth]{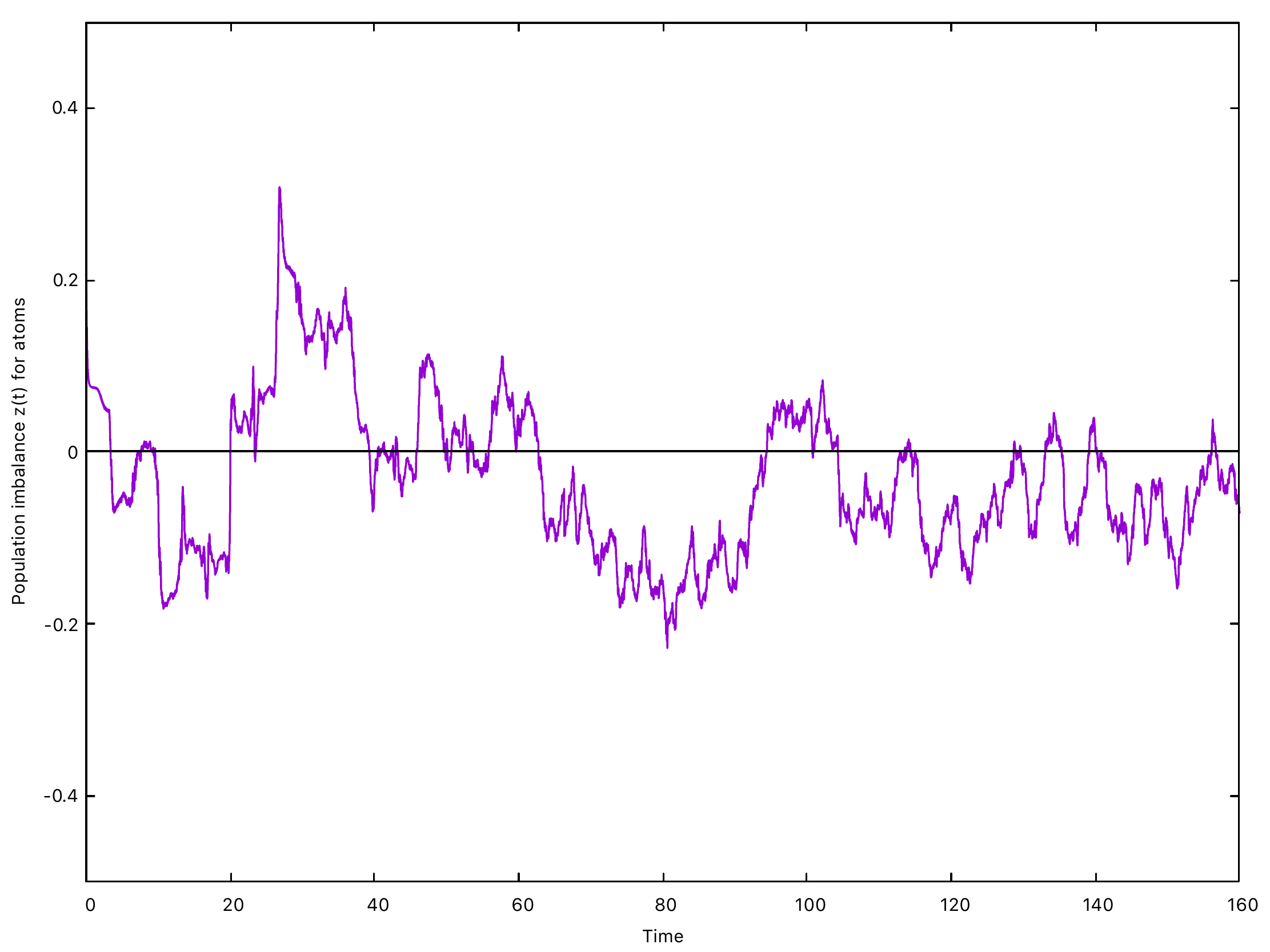}
{\caption*{(a)}}
\end{subfigure}
\hfill
\begin{subfigure}[t]{0.45\textwidth}
\includegraphics[width=\textwidth]{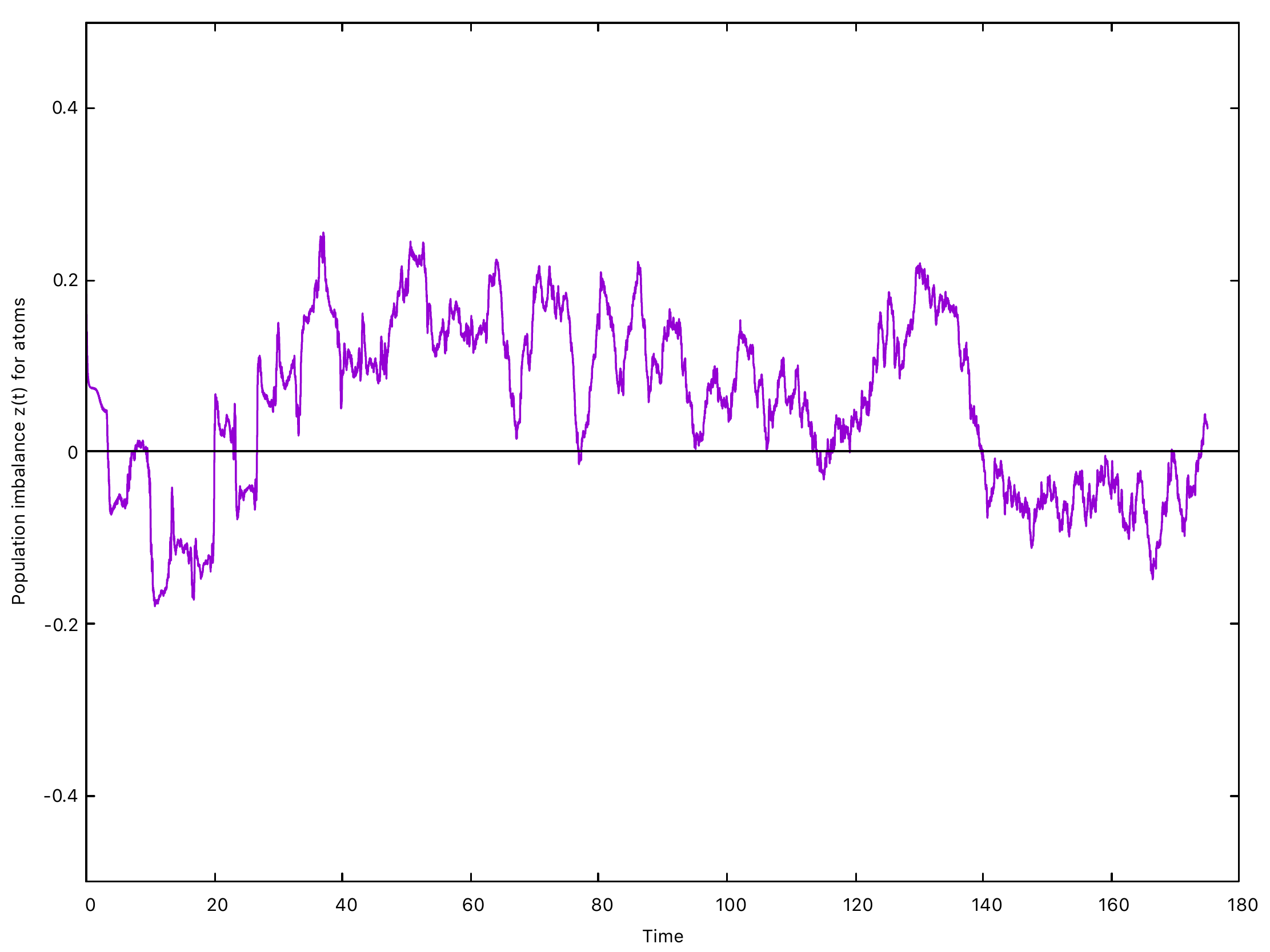}
{\caption*{(b)}}
\end{subfigure}

{\caption*{Figure 5: Population imbalance for atoms for (a) $\chi\neq 0$ and $ a_{dd}=0$ and (b) for $\chi=0$ and $ a_{dd}=0$}}
\end{figure}

However for molecules the (lower panels of fig.4(b)) the transient
effect of self trapping in the left well in absence of coherent
coupling ($a_{dd}=0$ and $\chi=0$) is not so prominent as that in
atoms. The molecular population in left well dominates over right
well only for short period of time t=125 to 140 approximately.
Otherwise it oscillates between two wells with the population in
the left well greater than that in the right well in most of the
time after $t=140$. This feature is demonstrated in the plot of
$z_m(t)$ in Fig.6(b). Similarly the effect of coherent coupling
(for $a_{dd}=0$ and $\chi\neq 0$) leading to transient
transmission to the right well of the molecules (in the lower
panel of Fig.4(a)) is not so prominent as in the case of atoms.
Here molecular population in right well dominates over that in the
left well only in the short period of times, the longest of which
is t=100 to 125 approximately at a stretch. This feature is
evident from the corresponding plot of $z_m(t)$ in Fig.6(a).
\begin{figure}
\begin{subfigure}[t]{0.45\textwidth}
\centering
\includegraphics[width=\textwidth]{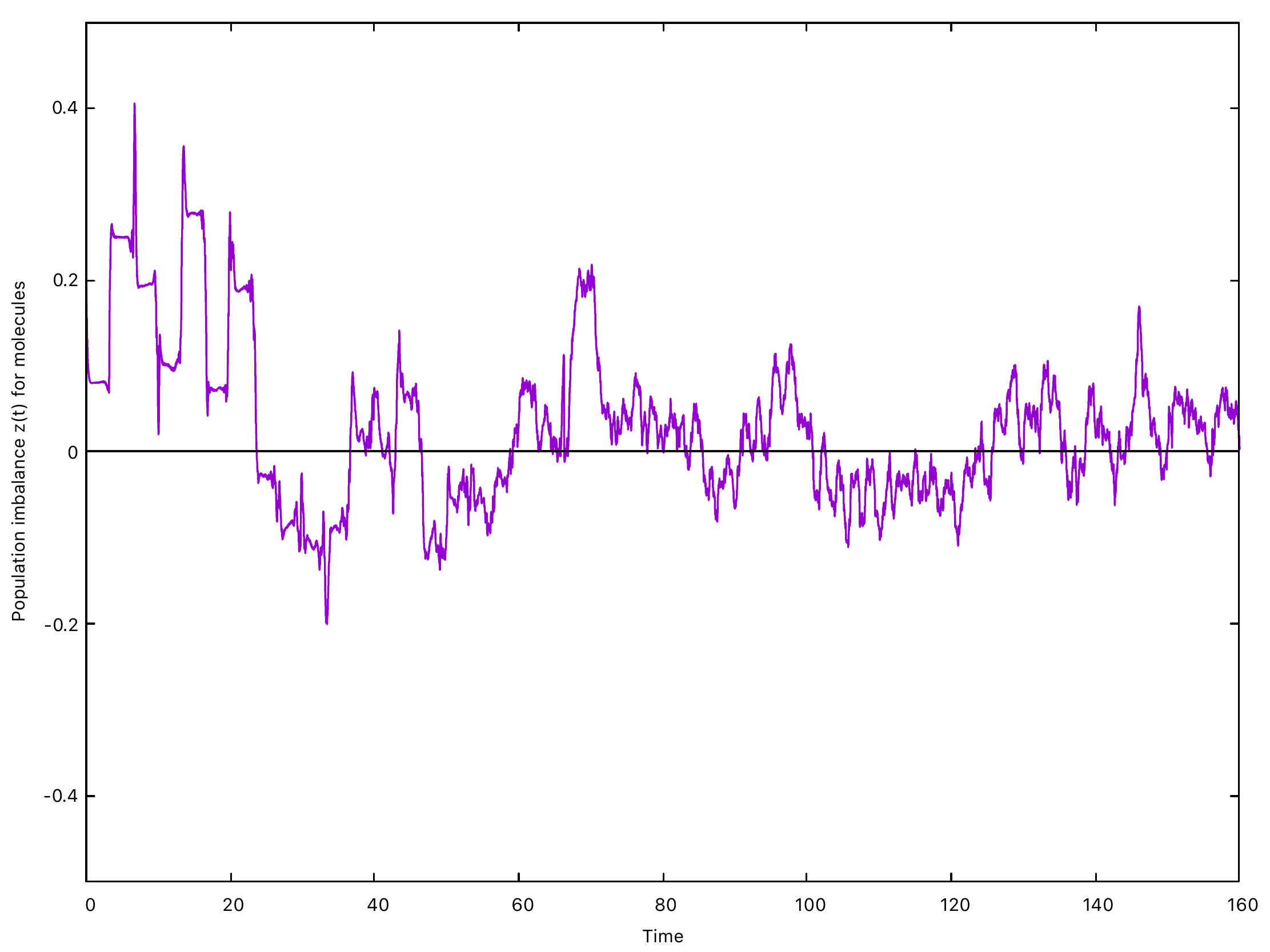}
{\caption*{(a)}}
\end{subfigure}
\hfill
\begin{subfigure}[t]{0.45\textwidth}
\includegraphics[width=\textwidth]{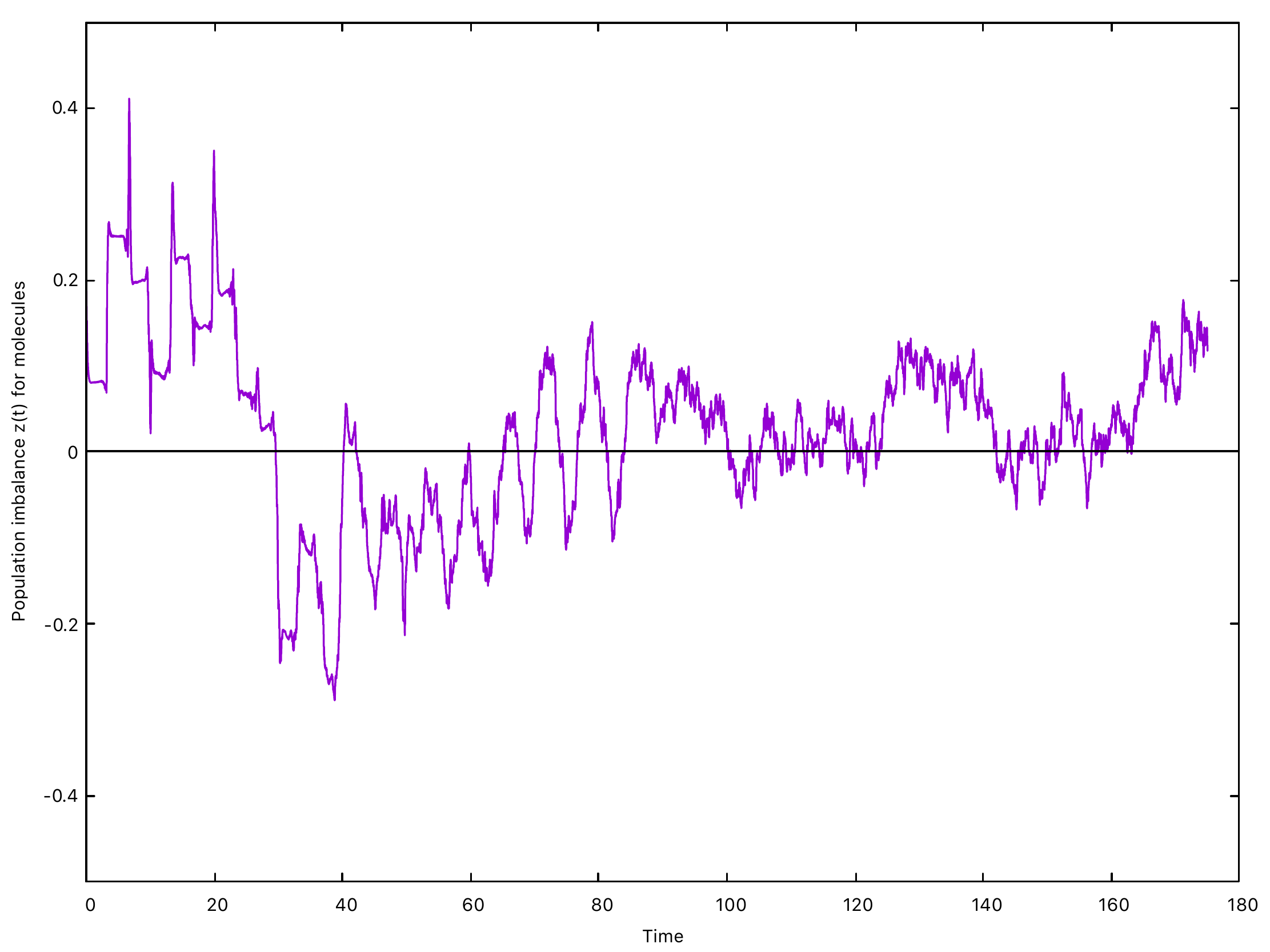}
{\caption*{(b)}}
\end{subfigure}

{\caption*{Figure 6: Population imbalance for molecules for (a) $\chi\neq 0$ and $ a_{dd}=0$ and (b) for $\chi=0$ and $ a_{dd}=0$}}
\end{figure}

Therefore from these figures it is found that the effect of
coherent coupling in absence of dipole-dipole interaction on
molecular dynamics is not so prominent as that in the case of
atoms. The dipole-dipole long range interaction and the coherent
coupling between atoms and molecules both are positive in this
calculation and both lead to the transient transmission of
population from the left well to the right well. The degree of
this transient transmission effect will depend on the strength of
these interactions which depends also on the number of particles
taking part in the dynamics. The dipole-dipole interaction is
proportional to the number of particles but the coherent coupling
term is proportional to the square root of the number of
particles. Hence in the absence of dipole-dipole interaction the
effective strength of interaction leading to transient
transmission is reduced. Moreover the number of molecules taking
part in the dynamics is much less than the number of atoms and
hence the transient transmission effect is much less effective in
case of molecules than that in case of atoms in absence of
dipole-dipole interaction.
\subsection{Effect of dipole-dipole interaction on the dynamics of population in presence and absence of coherent coupling:}
The dipole-dipole interaction which is a long range interaction,
is expected to facilitate the transmission of atomic and molecular
population from the left well to the right well when it is chosen
to be positive in nature. In absence of coherent coupling
$(\chi=0)$ the effect of dipole-dipole interaction on the dynamics
is demonstrated in figures 1(b) and 4(b). Comparison of these two
figures shows that the long duration of self trapping of atoms in
left well in Fig.4(b) is shortened due to the effect of long range
dipole-dipole interaction which prefers to transmit the atoms from
left to right well. Comparison of dynamical behaviour of molecules
(lower panel of figures 1(b) and 4(b)) shows transient
transmission effect is present (in the lower  panel of Fig.1(b))
which is the signature of presence of dipole-dipole interaction
for ($\chi=0$).

 Moreover in presence of coherent coupling the transient transmission of atomic population from left to right well is present both in the presence and absence of dipole-dipole interaction as shown in Fig.1(a) and Fig.4(a) respectively. However duration of transmission is reduced in absence of dipole-dipole interaction. Similarly for molecules the transient transmission effect is prominent in presence of dipole-dipole interaction (lower panel of figure 1(a)).
\subsection{Effect of coherent coupling on the dynamics of atomic density profiles:}
To investigate the dynamical behaviour of density profile of atoms
in presence and absence of coherent coupling, we have plotted the
atomic density as a function of z in the time range of t=76 to
120.
\begin{figure}
\centering
\begin{subfigure}[b]{0.325\textwidth}
\centering
\includegraphics[width=\textwidth]{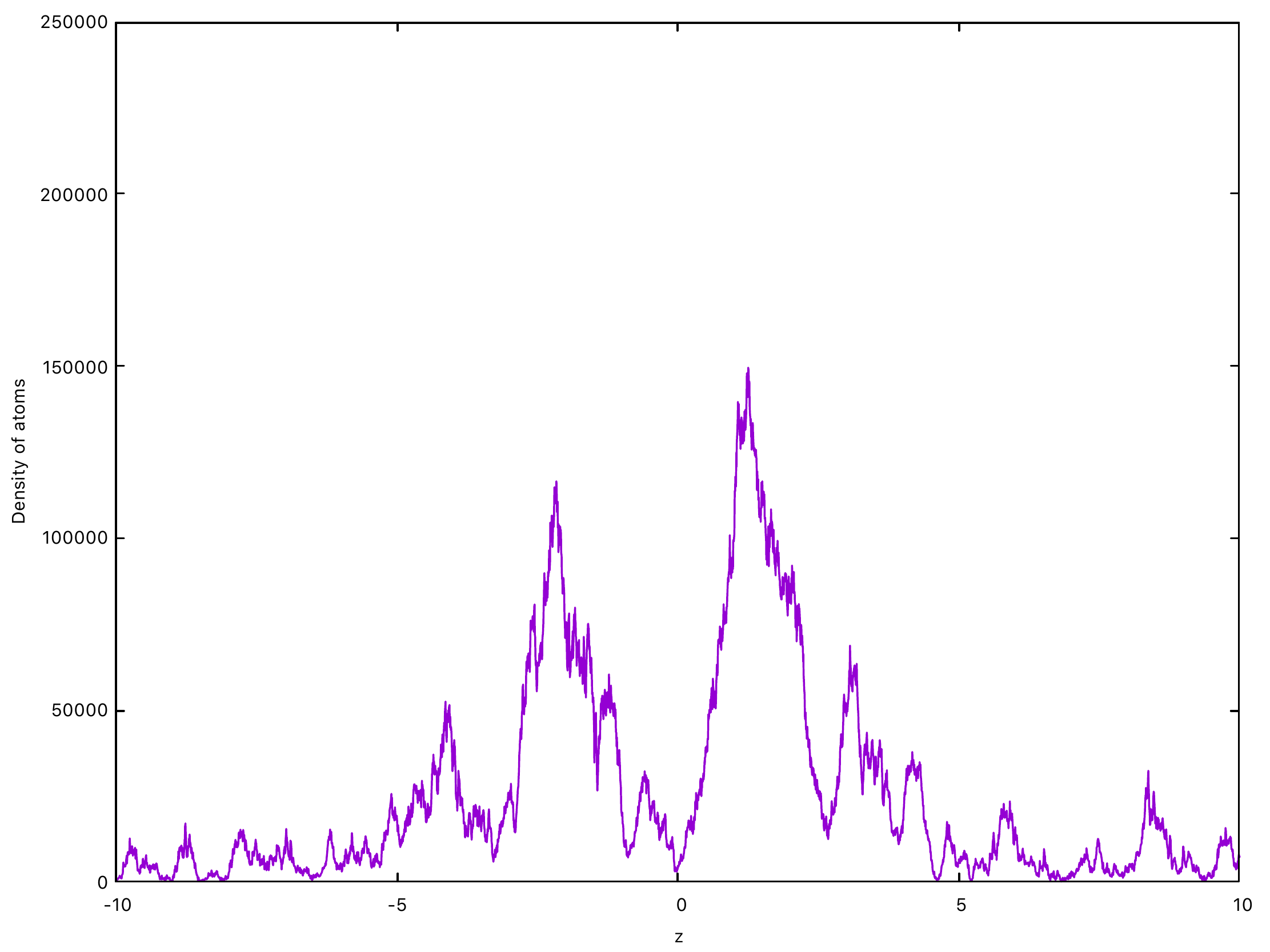} 
{\caption*{(a)}} 
\end{subfigure}
\hfill
\begin{subfigure}[b]{0.325\textwidth}
\centering
\includegraphics[width=\textwidth]{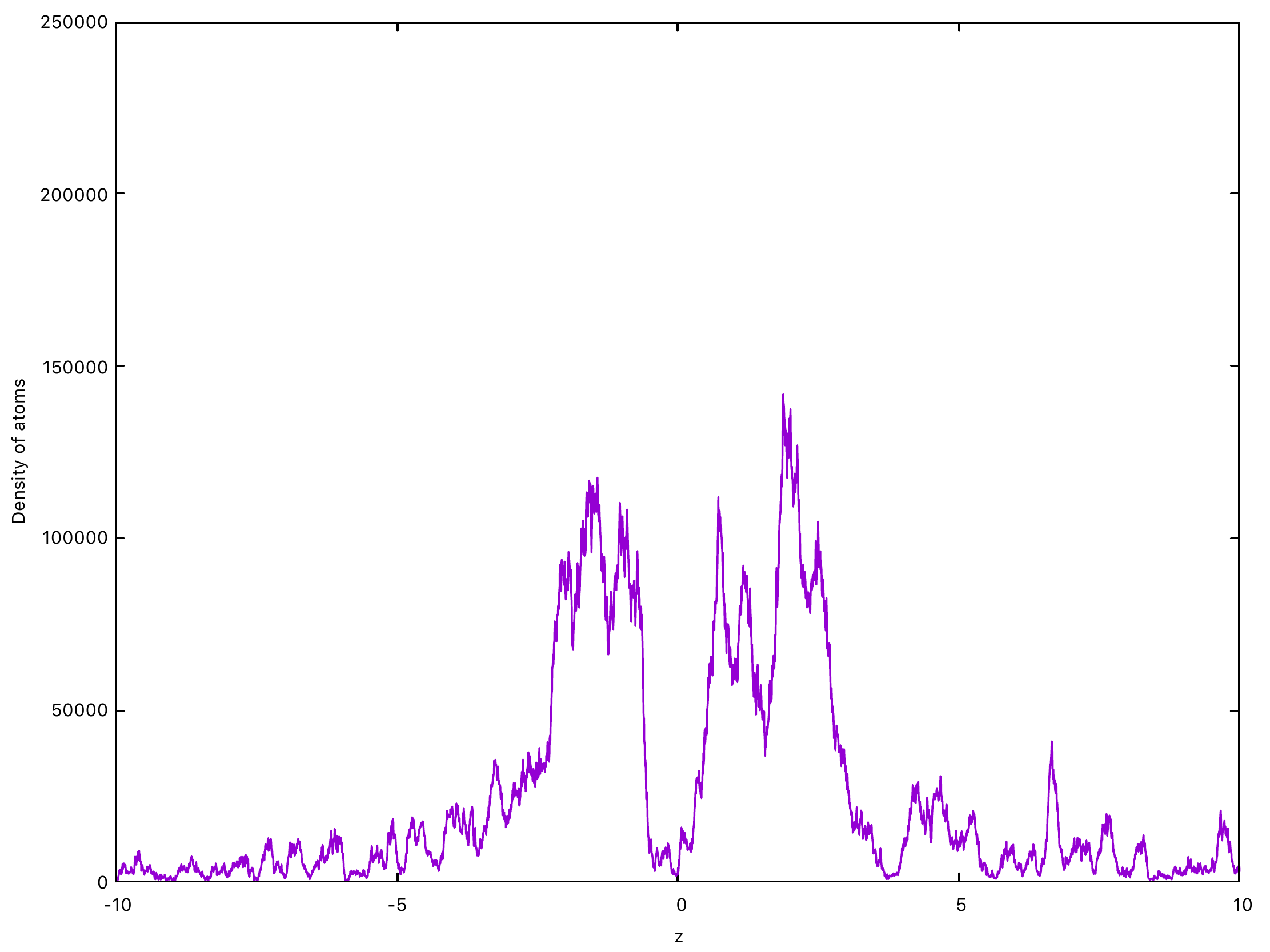} 
{\caption*{(b)}}
\end{subfigure}
\hfill
\begin{subfigure}[b]{0.325\textwidth}
\centering
\includegraphics[width=\textwidth]{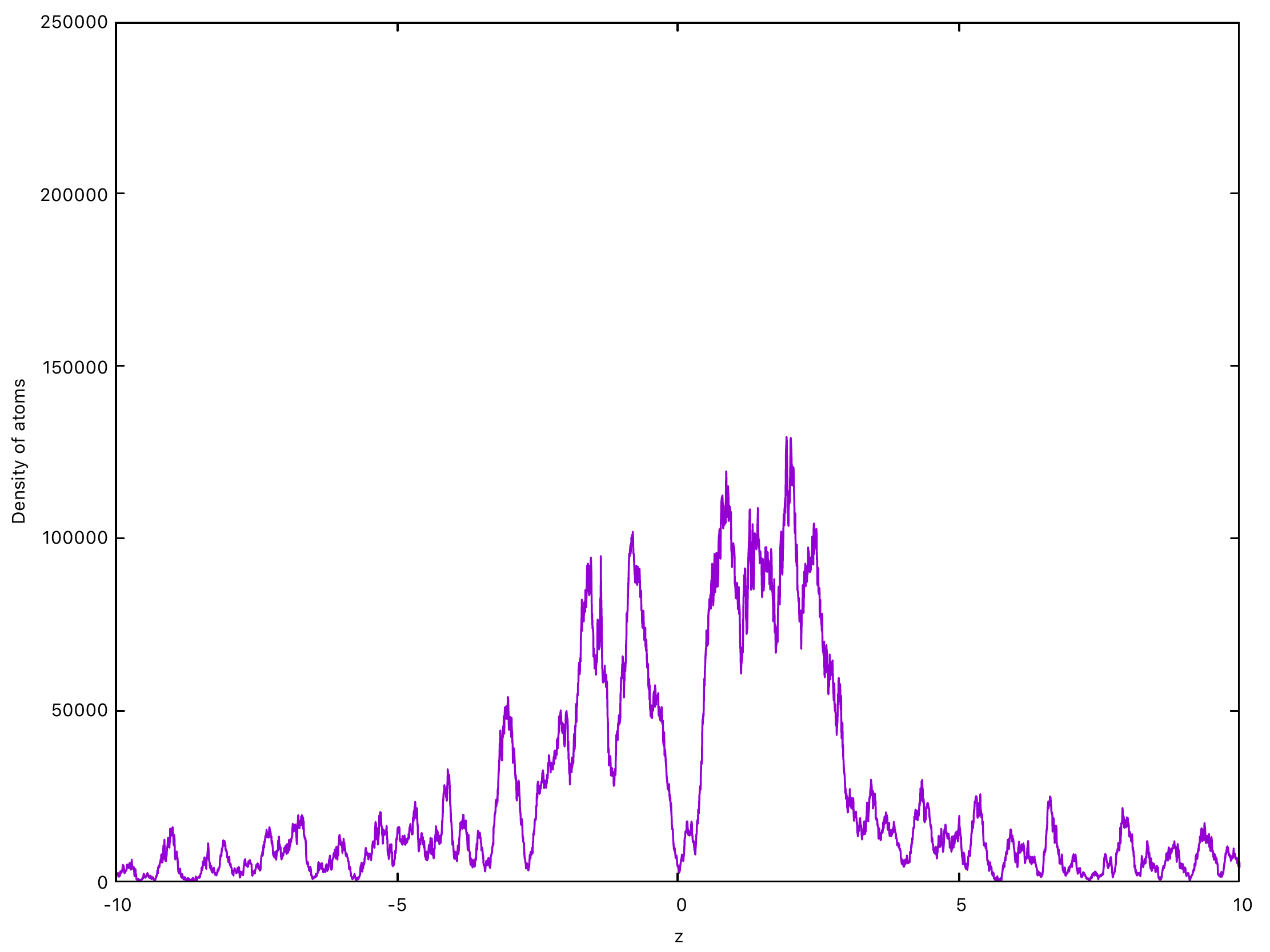} 
{\caption*{(c)}} 
\end{subfigure}

\begin{subfigure}[t]{0.325\textwidth}
\centering
\includegraphics[width=\textwidth]{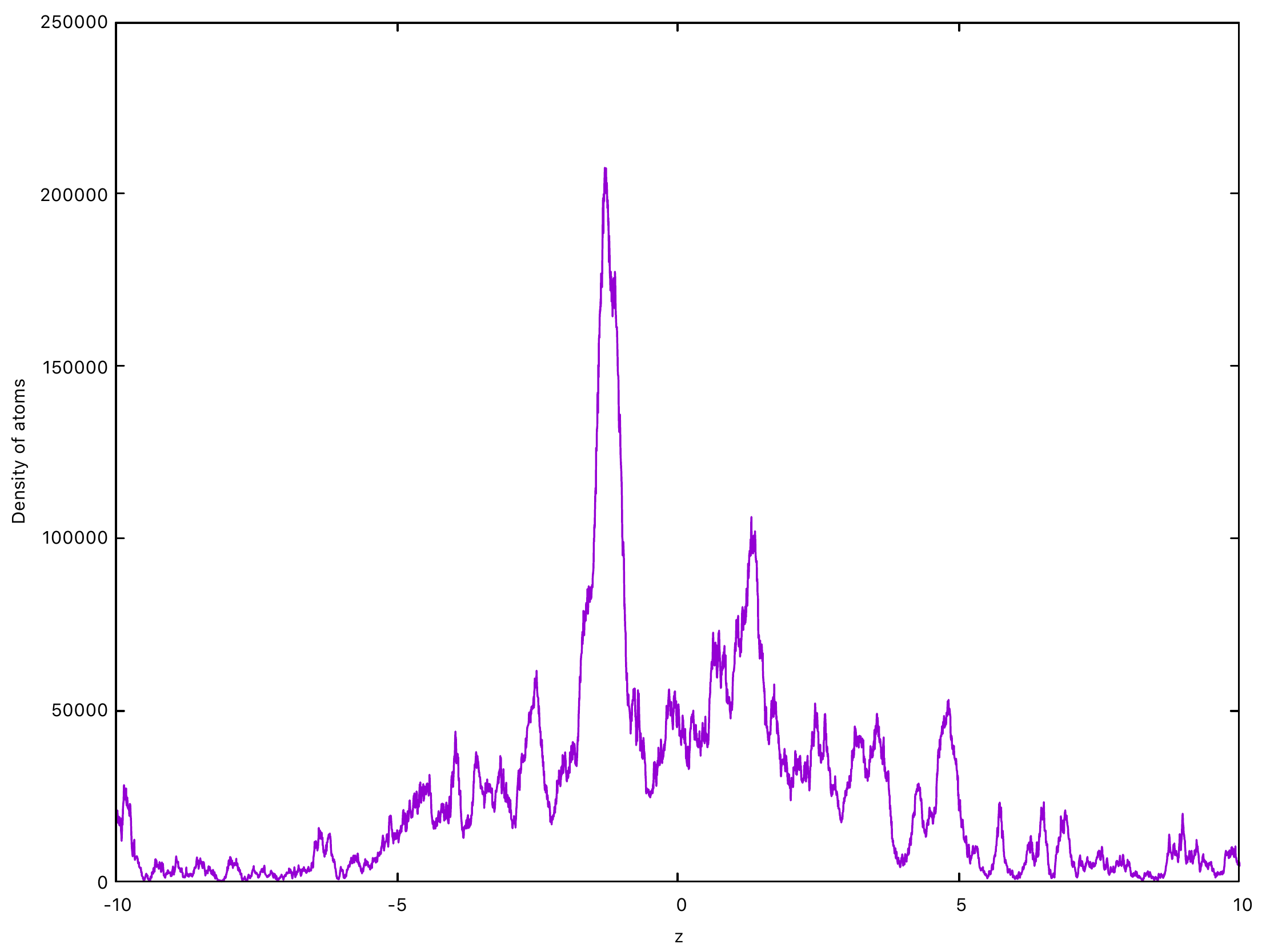} 
{\caption*{(d)}} 
\end{subfigure}
\hfill
\begin{subfigure}[t]{0.325\textwidth}
\centering
\includegraphics[width=\textwidth]{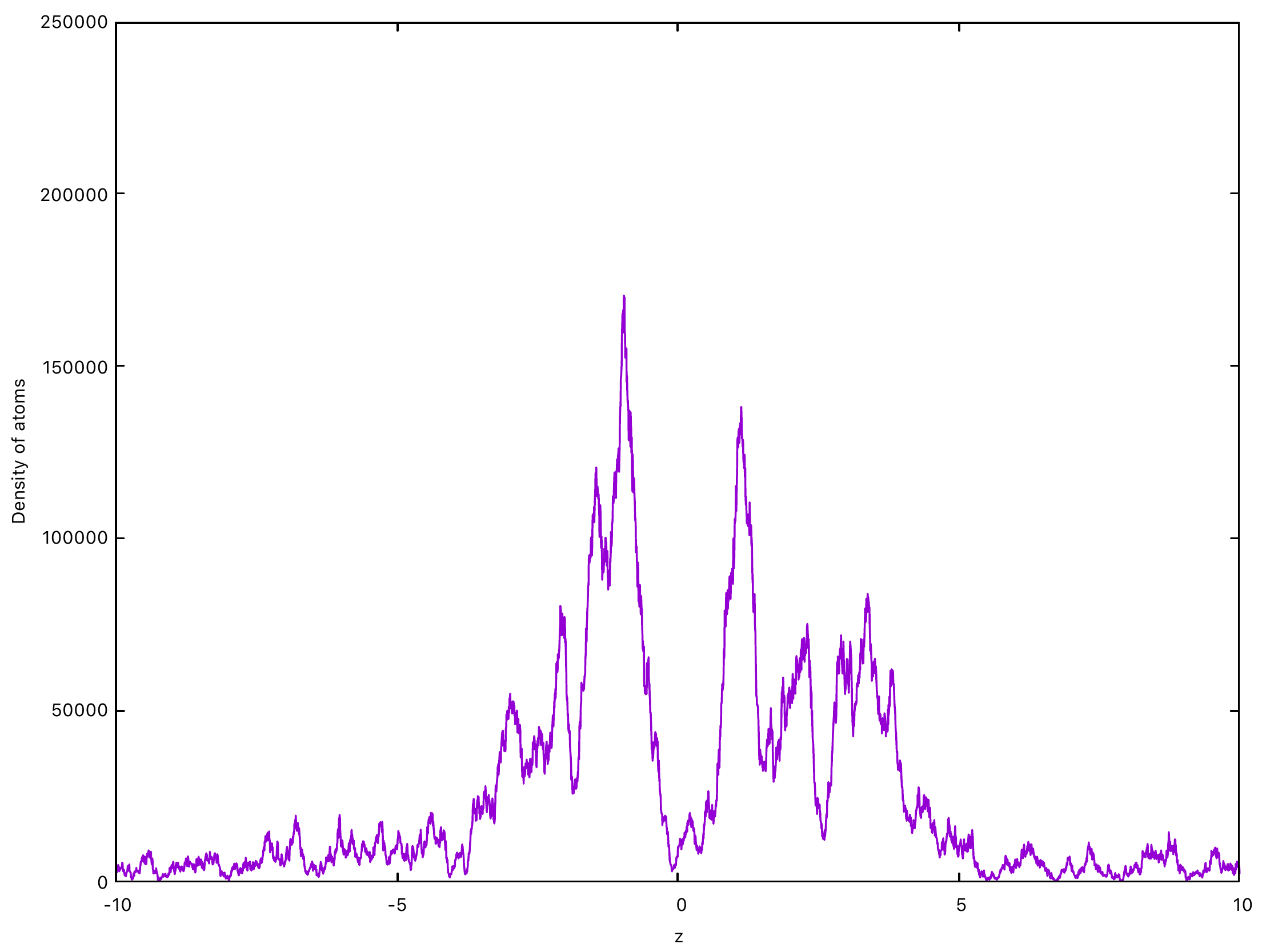} 
{\caption*{(e)}} 
\end{subfigure}
\hfill
\begin{subfigure}[t]{0.325\textwidth}
\centering
\includegraphics[width=\textwidth]{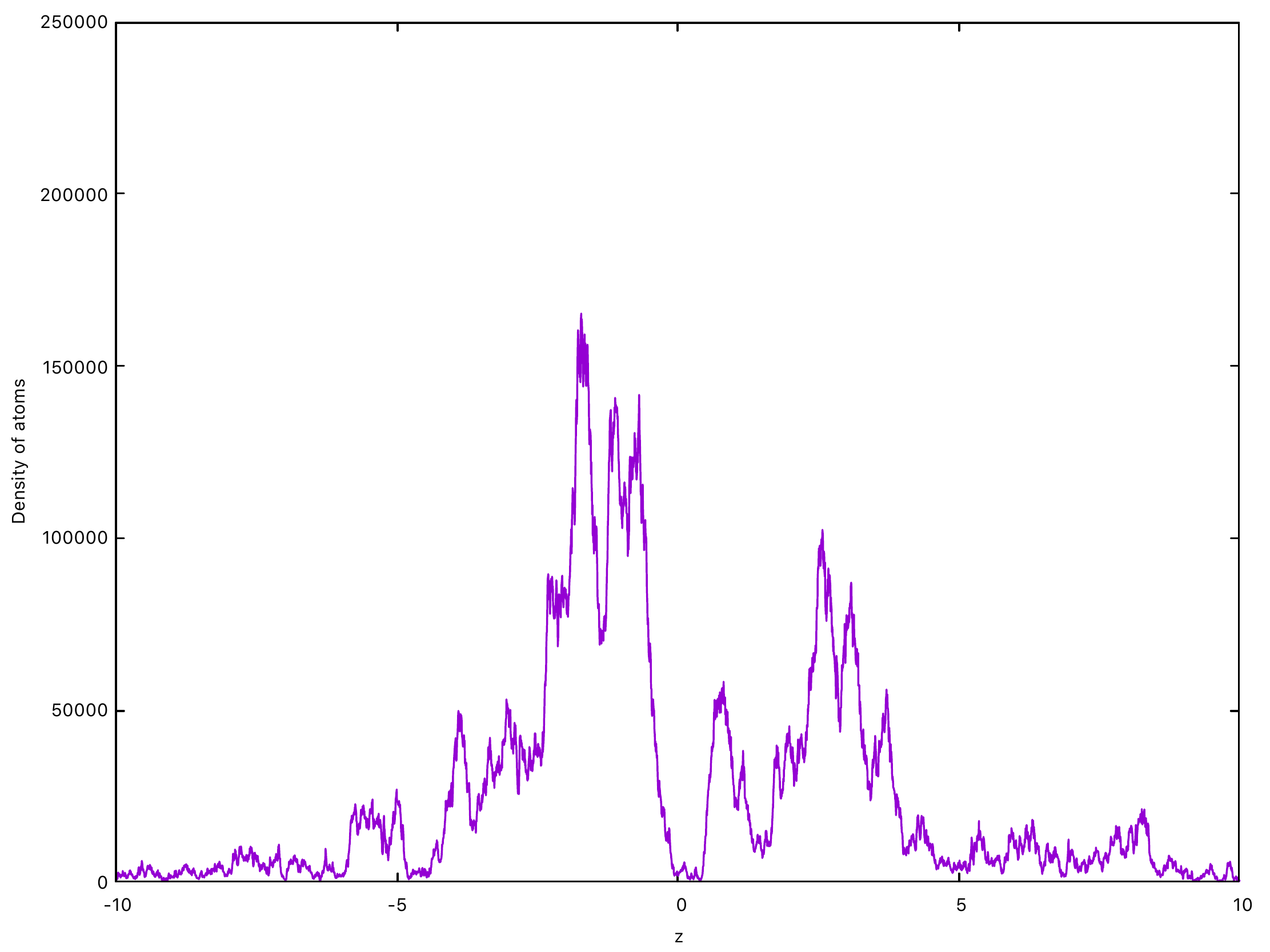} 
{\caption*{(f)}} 
 \end{subfigure}
 
{\caption*{Figure 7: Effect of coherent coupling on the dynamics of atomic density for $a_{dd}=.75a$. Upper panel in presence of coherent coupling $\chi$ at times (a) t=80, (b) t=101,(c) t=118  and lower panel in absence of it at times (d) t=80, (e) t=100 and (f) t=120.}}
\end{figure}

\begin{figure}
\centering
\begin{subfigure}[b]{0.325\textwidth}
\centering
\includegraphics[width=\textwidth]{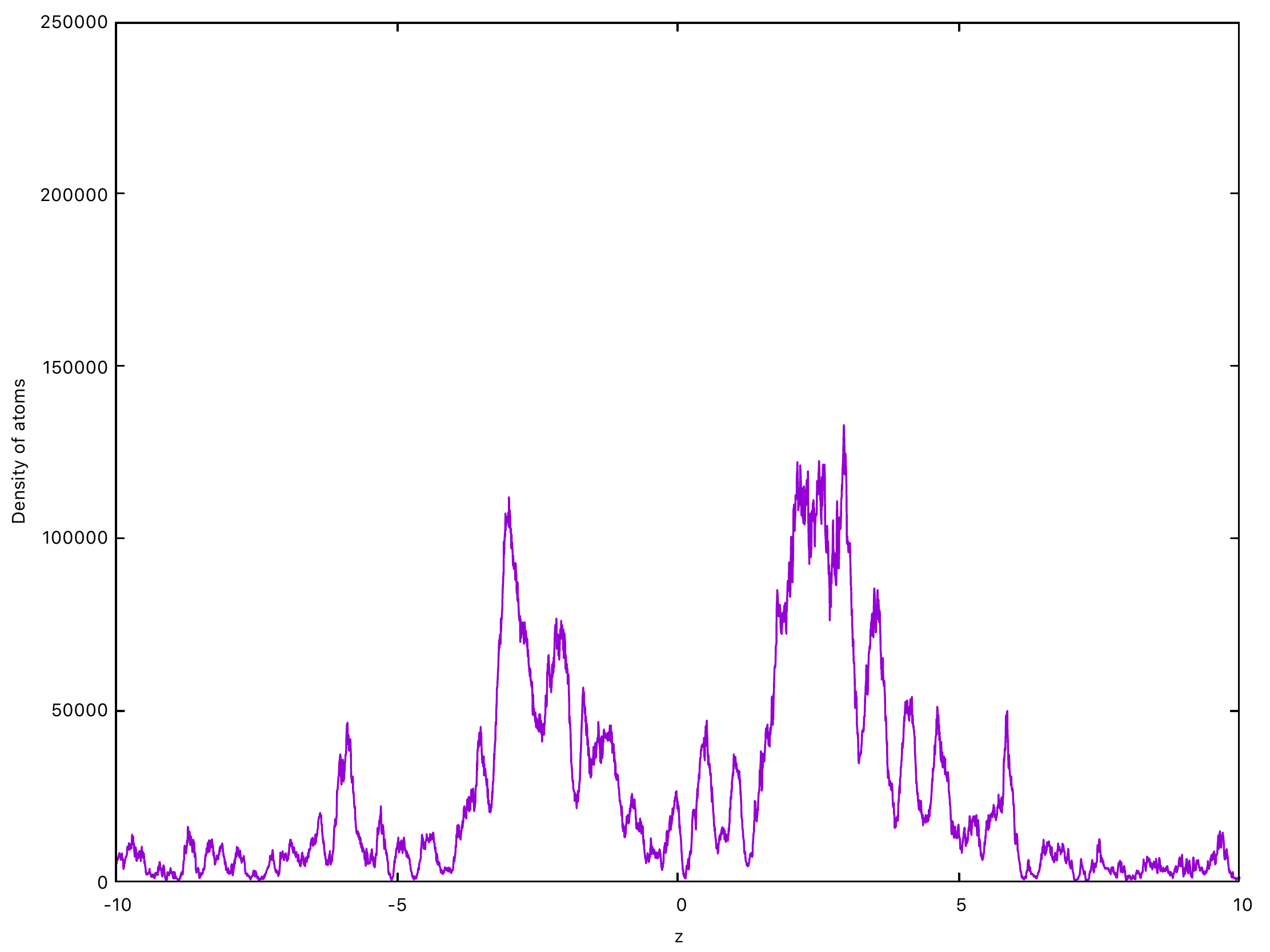} 
{\caption*{(a)}} 
\end{subfigure}
\hfill
\begin{subfigure}[b]{0.325\textwidth}
\centering
\includegraphics[width=\textwidth]{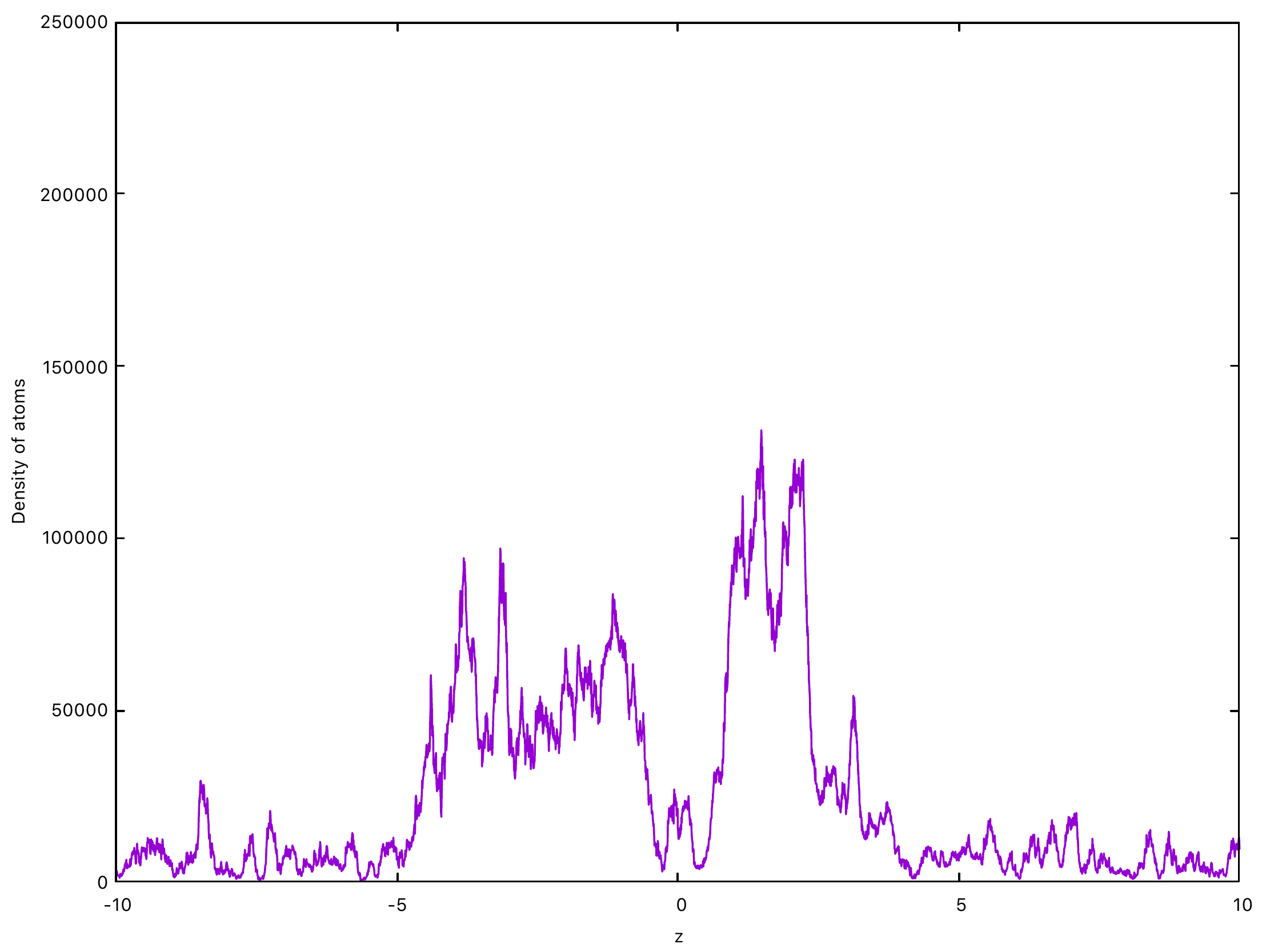} 
{\caption*{(b)}}
\end{subfigure}
\hfill
\begin{subfigure}[b]{0.325\textwidth}
\centering
\includegraphics[width=\textwidth]{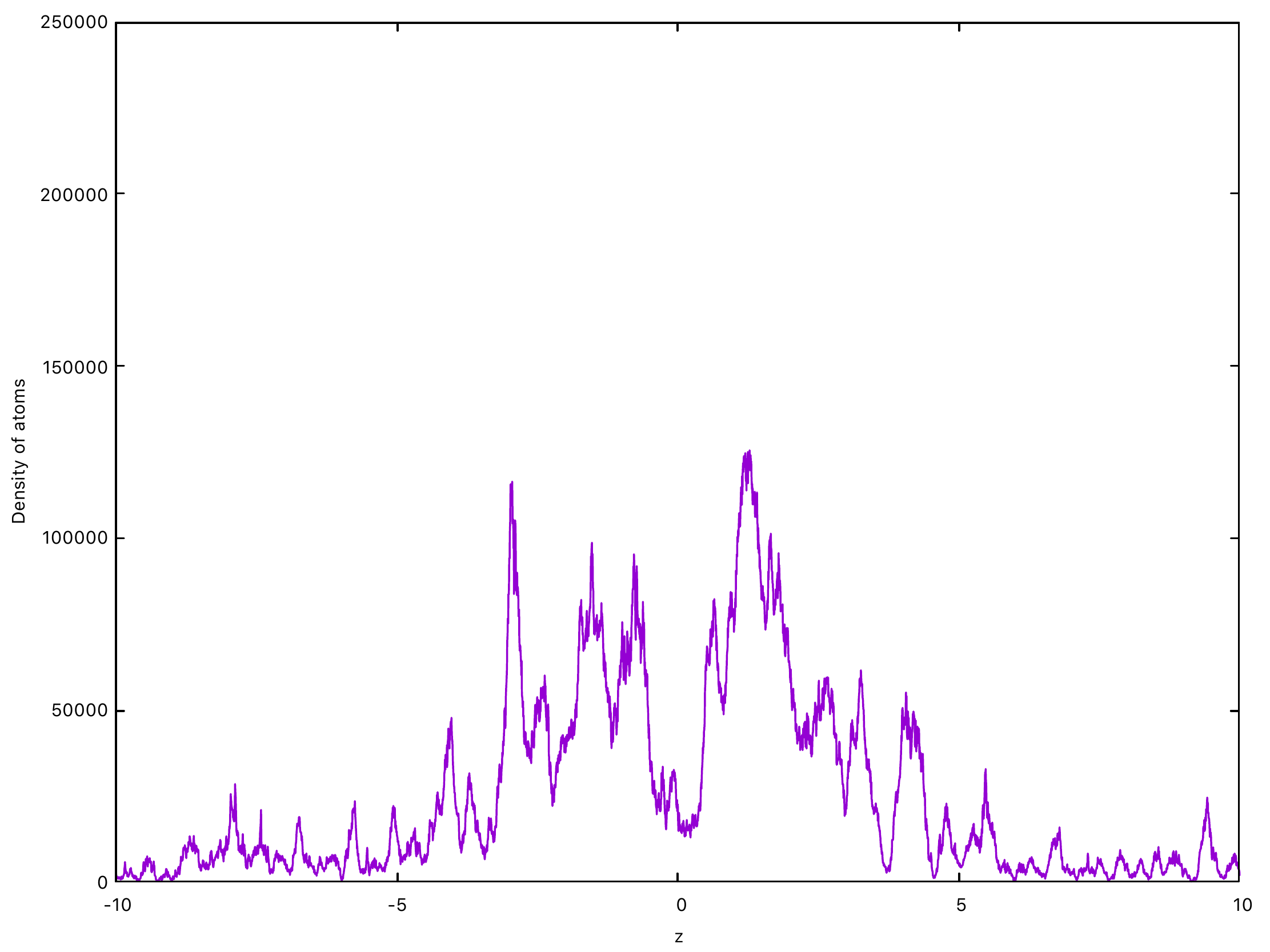} 
{\caption*{(c)}} 
\end{subfigure}

\begin{subfigure}[t]{0.325\textwidth}
\centering
\includegraphics[width=\textwidth]{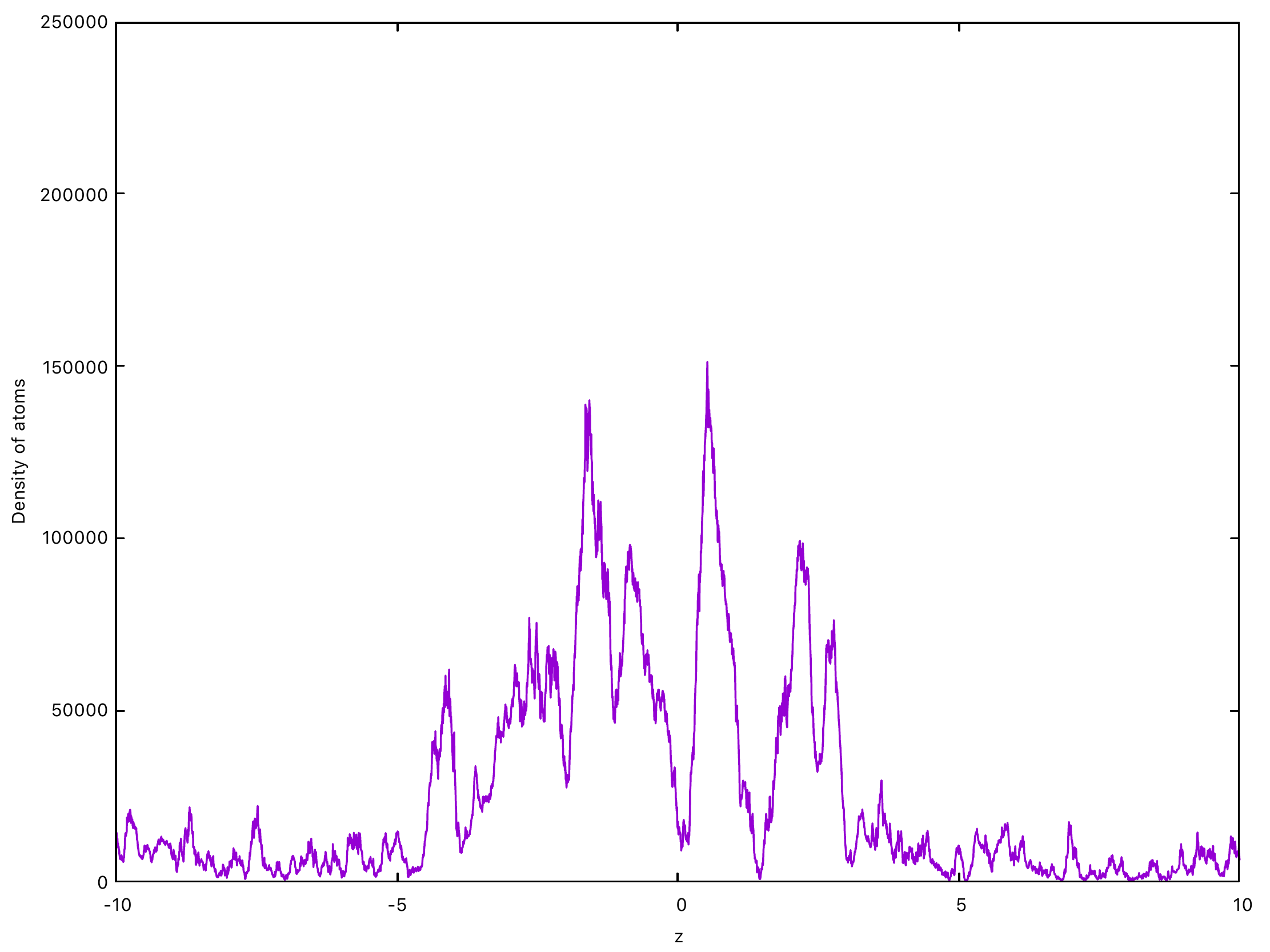} 
{\caption*{(d)}} 
\end{subfigure}
\hfill
\begin{subfigure}[t]{0.325\textwidth}
\centering
\includegraphics[width=\textwidth]{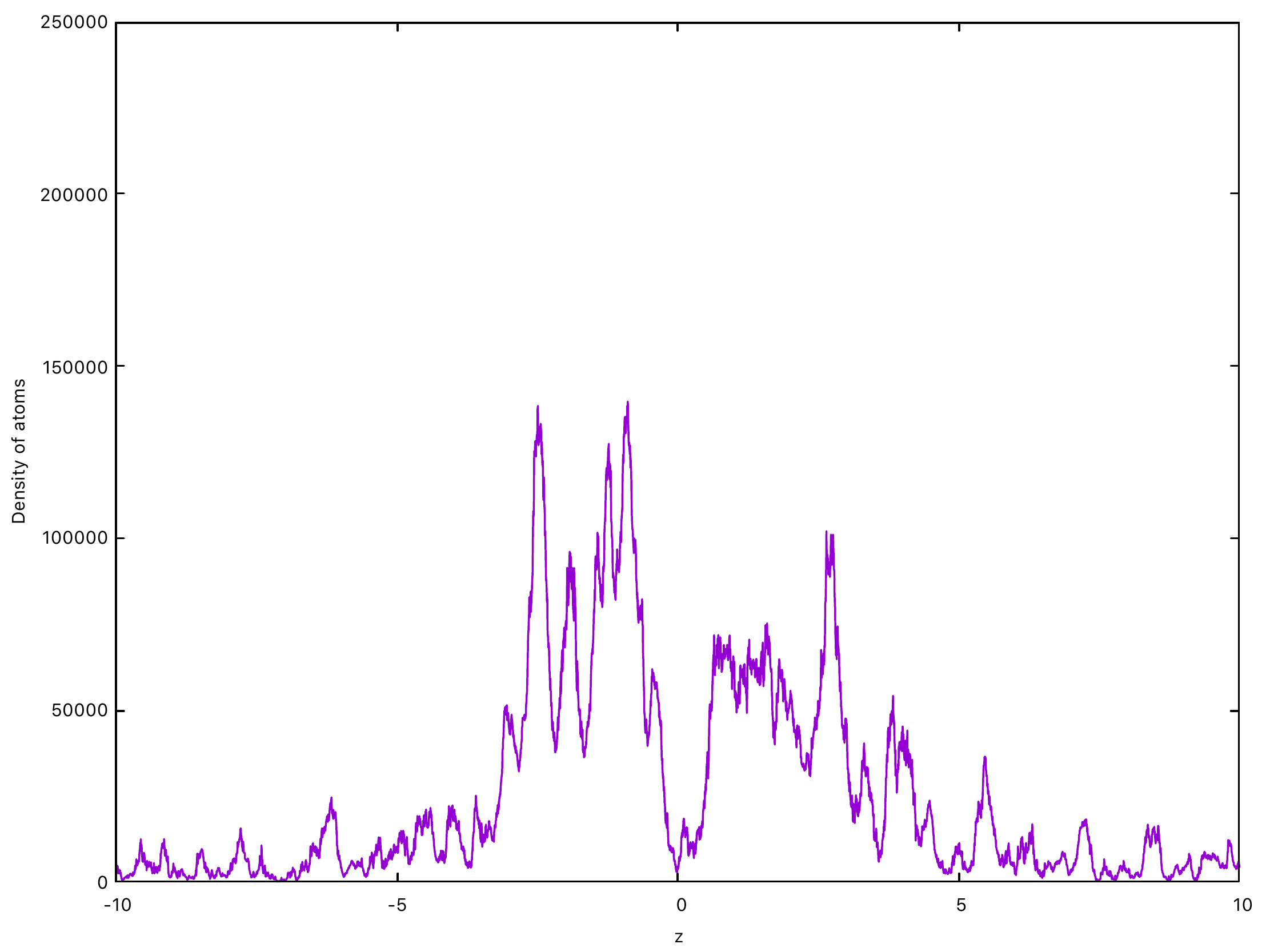} 
{\caption*{(e)}} 
\end{subfigure}
\hfill
\begin{subfigure}[t]{0.325\textwidth}
\centering
\includegraphics[width=\textwidth]{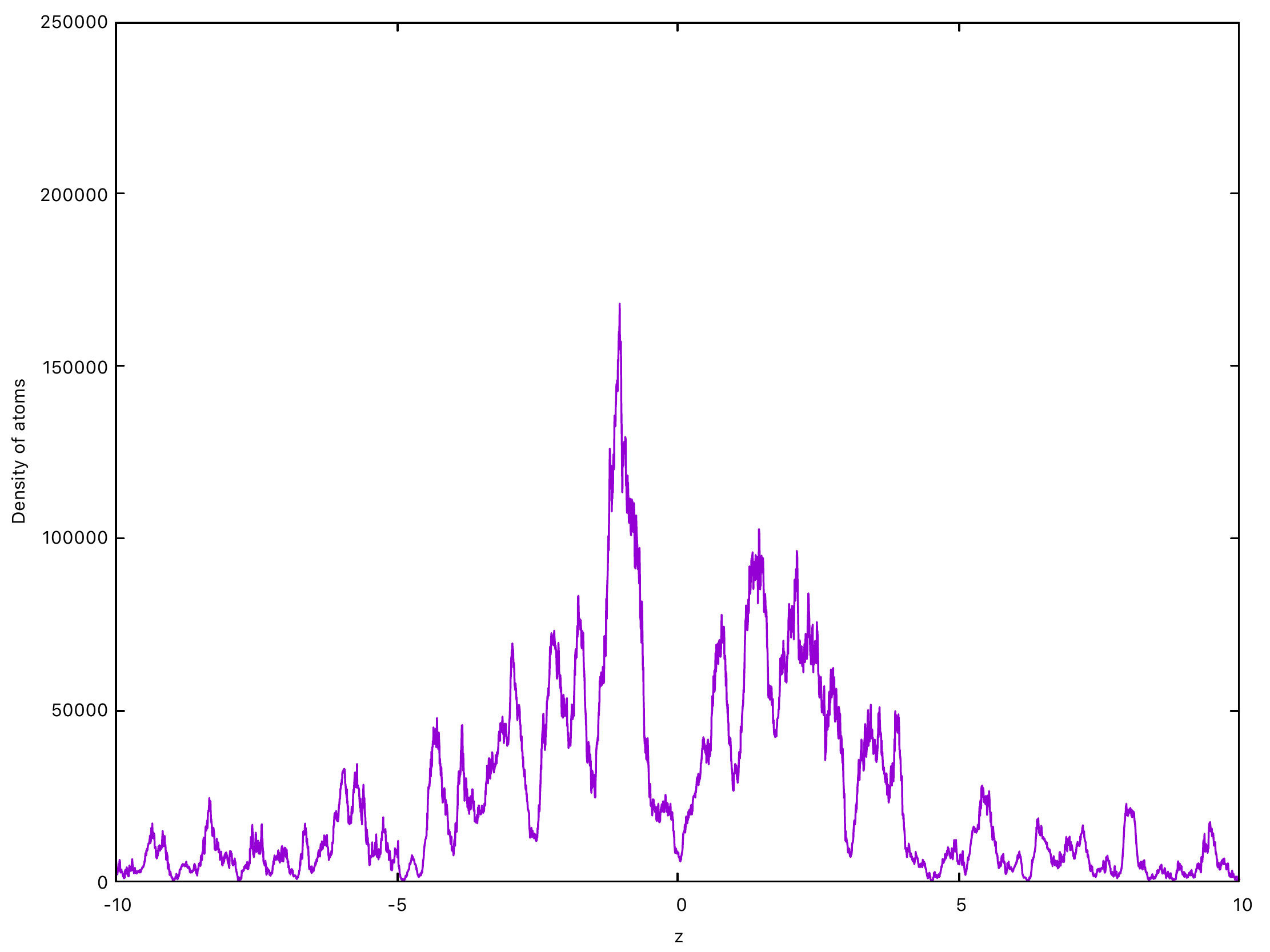} 
{\caption*{(f)}} 
 \end{subfigure}
 
{\caption*{Figure 8: Effect of coherent coupling on the dynamics of atomic density for $a_{dd}=0$. Upper panel in presence of
    coherent coupling $\chi$ at times (a) t=80, (b) t=101,(c) t=118  and lower panel in absence of it at times (d) t=80, (e) t=100 and (f) t=120.}}
\end{figure}

Figure 7 compares the nature of atomic density profile in two
cases for $a_{dd} \neq 0$ and  $\chi \neq 0$ at times t=80, 101
and 118 (in the upper panel) and $a_{dd} \neq 0$ and $\chi=0$ at
times t=80, 100 and 120 (in the lower panel). It is found that the
contribution of density of atoms is greater in right well than
that in the left well for non-zero coherent coupling (upper
panel). Whereas the density in left well is greater than that in
right well for $\chi=0$ (lower panel).

Similarly Fig.8 shows the dynamical behaviour of atomic density
profile for $a_{dd}=0$ and $\chi \neq 0$ at times t=76, 96 and 120
(in the upper panel) and $a_{dd}=0$ and $\chi=0$ at times t=80,
100 and 120 (in the lower panel). As in the presence of
dipole-dipole interaction shown in Fig.7, the effect of coherent
coupling (i.e. $\chi \neq 0$)  leads to more contribution in right
well than that in the left well even in absence of dipole-dipole
interaction. Whereas for $\chi=0$ the atomic density has more
contribution in left well compared to that in the right well.
Therefore these two figures show that in the time range of t=76 to
120 coherent coupling keeps its signature on the dynamics of
atomic density profiles leading to more contribution in the right
well than that in the left well and this effect persists both in
presence and absence of dipole-dipole interaction.

\section{Conclusion:} 
In the present model calculation we have 
studied numerically the dynamics of
coupled dipolar atomic and molecular BECs trapped in a double well
potential. The effect of coherent coupling between atoms and
molecules (through  magnetic Feshbach coupling)  on the dynamics
of atomic and molecular population between two wells has been
investigated in presence and absence of long range dipole-dipole
interaction. We have solved four time-dependent coupled GP like
equations two for left well and two for right well (for atoms and
molecules) which includes both the long range dipole-dipole
interaction and short range contact interaction. These equations
have been derived from the variation of energy functional as it is
done for the derivation of GP equations. Then the three
dimensional equations have been reduced to single dimension (in
the axial direction) by considering the strong confinement in the
radial direction to study the dynamics of population in the double
well which is considered to be in the axial direction. To
demonstrate the effect of coherent coupling and the dipole-dipole
interaction the evolution of total number of atoms and molecules
in the left and in the right well has been shown. Corresponding
population imbalance as a function of time has been studied to
support the results. In the present work dipole-dipole interaction
has been chosen to be positive and since it is long range in
nature, it leads to transient transmission of atomic and molecular
populations from the left to the right well. Moreover the coherent
coupling inherent in this coupled system is also positive and hence 
intensifies the transient transmission of population from the left 
to the right well. As a result the absence of anyone of these two 
positive interactions leads to reduction of transient transmission 
of population. It is shown that in the absence of these two
positive interactions the self trapping of population in the left 
well is prominent for a long duration which is disturbed in presence 
of anyone of these positive interactions either the long range 
dipole-dipole interaction or the coherent interaction between 
atoms and molecules. These effects are also demonstrated by the 
dynamics of population imbalance for atoms and molecules in each case.
The effect of coherent coupling on the density profile of atoms has 
been demonstrated by plotting the atomic density as a function of z 
at three instant of time from t=76 to 120 (t in the units of 
$\omega_z^{-1}$). It is shown that the presence of coherent 
coupling leads to concentration of population in the right well more than
that in the left well both in presence and absence of dipole-dipole interaction.\\

{\bf{Acknowledgement}}:
We thank  Bimalendu Deb, for useful discussions and Satrajit Adhikari for providing computational facility in cluster computer installed in his laboratory.

\end{document}